\documentclass[aps, prl,
 reprint,
 superscriptaddress,
 citeautoscript,
 amsmath,amssymb,
 tightenlines
]{revtex4-2}

\usepackage{graphicx}
\usepackage{dcolumn}
\usepackage{bm}
\usepackage{lipsum}
\usepackage{booktabs}

\usepackage{hyperref}
\usepackage[capitalize]{cleveref}

\begin{document}
\setcitestyle{super}
\preprint{APS/123-QED}

\title{Non-Thermal Aging of Supercooled Liquids in Optical Cavities}

\author{Muhammad R. Hasyim}
\affiliation{Department of Chemistry, New York University, New York, NY 10003}
\affiliation{The Simons Center for Computational Physical Chemistry, New York University, New York, NY 10003}

\author{Arianna Damiani}
\affiliation{Department of Chemistry, New York University, New York, NY 10003}

\author{Norah M. Hoffmann}
\email{norah.hoffmann@nyu.edu}
\affiliation{Department of Chemistry, New York University, New York, NY 10003}
\affiliation{The Simons Center for Computational Physical Chemistry, New York University, New York, NY 10003}
\affiliation{Department of Physics, New York University, New York, NY 10003}

\date{\today}

\begin{abstract}

\noindent Aging is a hallmark of disordered materials such as glasses, plastics, and pharmaceuticals, where it often limits long-term stability and performance. In practice, aging is controlled through global parameters like temperature or pressure, which act uniformly on the entire system. Here we introduce a fundamentally different approach, using light confined in optical cavities as a precise and selective tool to guide aging dynamics. We show that a supercooled liquid coupled to an optical cavity undergoes \emph{non-thermal aging}, where aging is induced by light without a thermal quench. Light selectively pumps fast vibrational modes while the bath temperature remains unchanged, reshaping the slow structural dynamics of the liquid. The cavity-coupled liquid thereby behaves as if it were structurally colder than its surroundings.
Exploiting this effective structural cooling together with the timescale separation, we introduce cavity configurational feedback ($\mathrm{C^2F}$) cooling, which uses cavity coupling to reach progressively lower structural temperatures. Our results establish a connection between glass physics and strong light–matter interactions and open a new route toward optical control of aging, glass formation, and nonequilibrium materials dynamics.

\end{abstract}

\maketitle

\begin{figure*}[t]
    \centering
    \includegraphics[width=0.95\linewidth]{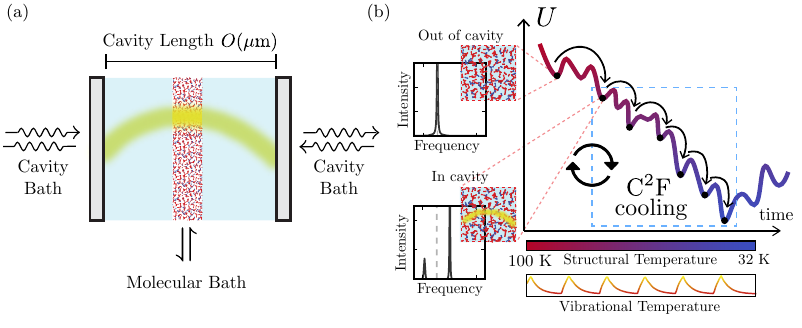}
    \caption{
    Overview of cavity-induced non-thermal aging in a supercooled liquid.
    (a) Schematic of a supercooled liquid in a Fabry--Pérot cavity with mirror spacing $L \sim O(\mu\mathrm{m})$.  
    The cavity mode (yellow line) selectively couples to molecular vibrations, allowing it to channel energy into specific intramolecular modes without uniformly heating the liquid.
    Both molecular and cavity subsystems exchange energy with their respective thermal bath at temperature $T$ (arrows).
    (b) Non-thermal effect on structural relaxation by a cavity. The top-left inset shows the vibrational spectrum out of cavity, while the bottom-left inset shows the spectrum in the cavity under strong coupling. The central schematic illustrates aging on the potential energy landscape $U$, where the system progressively explores deeper energy basins over time. Out of cavity, the supercooled liquid is at equilibrium ($T = 100$~K). Upon entering the strong light--matter coupling regime, evidenced by Rabi splitting in the IR spectrum (bottom-left inset), the system is driven into deeper basins corresponding to lower structural fictive temperatures (color bar, red to blue). C$^2$F cooling repeatedly applies this mechanism by switching the cavity on and off (blue dashed box) to reach progressively lower structural fictive temperature. A feedback loop maintains the bath temperature approximately equal to the structural fictive temperature to preserve structural equilibrium during the protocol. The faster vibrational temperature oscillates in time according to the feedback cycle.}
    \label{fig:fig1}
\end{figure*}

Many materials that underpin modern technology, such as glasses, polymers, batteries, pharmaceuticals, and solar cells, are structurally disordered.\cite{berthier2011theoretical,biroli2013perspective}
They are inherently out of equilibrium and slowly evolve toward more stable configurations; a process known as physical aging.\cite{struik1978physical,angell2000relaxation,kauzmann1948nature} 
Since aging proceeds on astronomical timescales, disordered materials never reach their optimal stability, making controlled acceleration of aging central to improving their long-term performance.\cite{rodriguez2022ultrastable,ediger2017perspective,chen2023geological} 
Current efforts to control aging rely on thermodynamic state variables such as temperature, pressure, or chemical composition,\cite{sun2012stability,dragic2018materials} but these operate with little specificity; temperature heats or cools the entire material, pressure acts on all degrees of freedom, and chemical modification alters the material itself.\cite{lu2013colloidal}

Here we take a fundamentally different approach, using electromagnetic fields confined in optical cavities to target specific molecular transitions at a chosen frequency.\cite{vahala2003optical,aspelmeyer2014cavity} 
When matter is placed inside the cavity, the cavity modes interact strongly with molecular degrees of freedom, enabling an efficient energy exchange between light and matter.\cite{cohen2024atom,cohen2024photons,ebbesen2023introduction,garciavidal2021manipulating,li2022molecular} 
Since light-matter energy exchange occurs on picosecond timescales whereas structural relaxation in supercooled liquids takes nanoseconds or longer, the strong interaction also creates a clear separation of timescales, allowing light to influence aging precisely and selectively rather than simply heating the material.\cite{thomas2016ground,Dunkelberger2022,schlawin2022cavity}

As an ideal testbed, our work explores strong coupling of a supercooled liquid in an optical cavity (\cref{fig:fig1}a). Supercooled liquids remain fluid and disordered at low temperatures without crystallizing, leading to extremely slow structural relaxation. We show that the cavity acts as a tunable non-thermal energy source, selectively pumping energy into dipole-active vibrations and inducing what we refer to as \emph{non-thermal aging} where aging is driven without changing the temperature. 
We find that non-thermal aging proceeds by lowering the structural fictive temperature, i.e., the equilibrium temperature at which the system would exhibit the same structural energy. Moreover, it can be predicted using only equilibrium relaxation data.
Since the fictive temperature can be controlled, we propose cavity configurational feedback ($\mathrm{C^2F}$) cooling (\cref{fig:fig1}b), a protocol that uses the cavity to access lower-temperature equilibrium structures and dynamical states that cannot be reached through purely thermal means. Our results suggest that strong light-matter interactions offer a new way to guide structural relaxation and nonequilibrium dynamics, opening a new direction for controlling aging in disordered materials.

\section*{Supercooled Liquid in a Cavity}

We simulate a model glass-forming liquid ($N=250$ molecules at $T=100$~K) coupled to a Fabry–Pérot cavity (\cref{fig:fig1}a). 
The molecular system is a dipole-active extension of the Kob–Andersen (KA) model;\cite{kob1995testing} a two-component Lennard-Jones fluid engineered to resist crystallization. The extension turns the Lennard-Jones particles into polar molecules with harmonic vibrational modes as well as partial charges that couple to the cavity field. 
The cavity mode frequency is set to $\omega_\mathrm{c} = 1560$~cm$^{-1}$, resonant with the intramolecular vibration of the red molecules, and two transparent spacers confine the liquid to the cavity's central region so that the mode's spatial structure can be neglected.\cite{gaber2015volume,li2020cavity,li2021cavity} 
Both subsystems exchange energy with external baths: a Langevin bath for the cavity mode and a Bussi–Parrinello bath\cite{bussi2007canonical,bussi2008stochastic} for the molecules, with the bath temperature held fixed to prevent cavity-induced heating. Note that all simulations use cavity molecular dynamics\cite{li2020cavity,li2021cavity} as implemented in \texttt{cavHOOMD-blue},\footnote{\url{https://github.com/muhammadhasyim/cav-hoomd}}\cite{anderson2020hoomd} with ensemble averages over $N_\mathrm{T}=500$ trajectories. 

\begin{figure*}[t]
    \centering
    \includegraphics[width=0.975\linewidth]{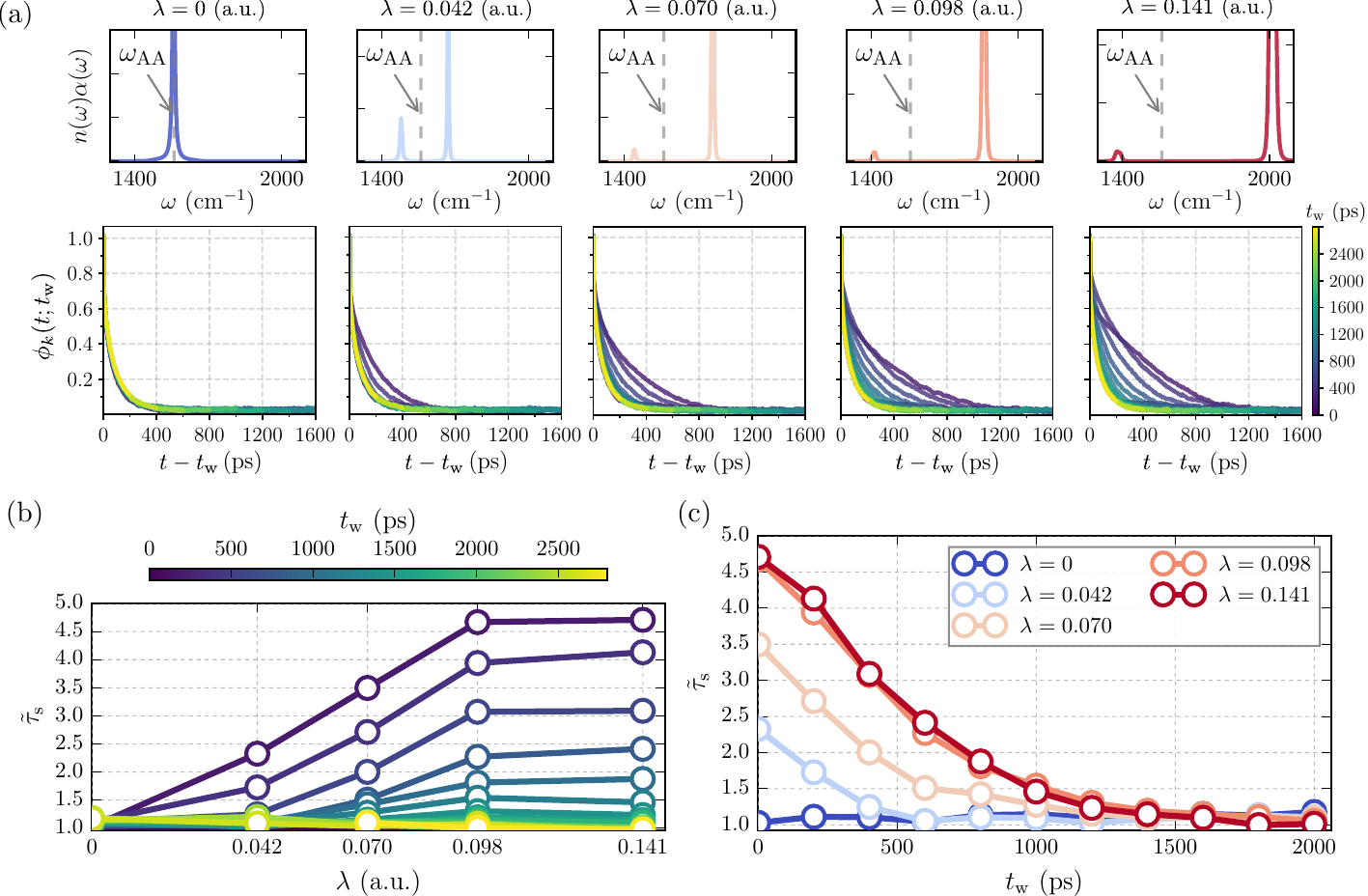}
    \caption{
    Cavity-induced slowdown of structural relaxation and non-thermal aging dynamics with $\omega_\mathrm{c} = 1560$~cm$^{-1}$ and $|\bm{k}| = 6.02$~a.u.
    (a) IR absorption spectra (top) and normalized ISFs $\phi_{\bm{k}}^{(\lambda)}(t;t_\mathrm{w})$ (bottom) for increasing coupling strength $\lambda$ (left to right) and waiting times $t_\mathrm{w}$ (color-coded, purple to yellow). Stronger coupling induces progressively slower relaxation with increasing dependence on $t_\mathrm{w}$, characteristic of aging.
    (b) Normalized structural relaxation time $\tilde{\tau}_\mathrm{s} = \tau_\mathrm{s}/\tau_{s,\lambda=0}$ versus $\lambda$ at different $t_\mathrm{w}$ (color-coded, purple to yellow), isolating the cavity-induced slowdown from intrinsic thermal aging.
    (c) $\tilde{\tau}_\mathrm{s}$ versus $t_\mathrm{w}$ for different coupling strengths, showing that the cavity-induced memory persists for up to $\sim 2.5$~ns.
    }
    \label{fig:figure2}
\end{figure*}

Strong coupling is verified from the IR spectra, computed via the dipole autocorrelation function\cite{tuckerman2023statistical} as shown in \Cref{fig:figure2}a (top panel). 
The IR spectrum from the cavity-free case (left, dark blue) to increasing light-matter coupling strengths, $\lambda = 0.042$--$0.141$~a.u. (left to right) shows how the original vibrational peak splits into upper and lower polariton branches, confirming the strong light-matter coupling regime.  Further details of the Hamiltonian, equations of motion, parameterization, and computational setup are provided in Methods and Supporting Information (SI)~Secs.~3 and~6.3.

\begin{figure*}[t]
    \centering
    \includegraphics[width=\linewidth]{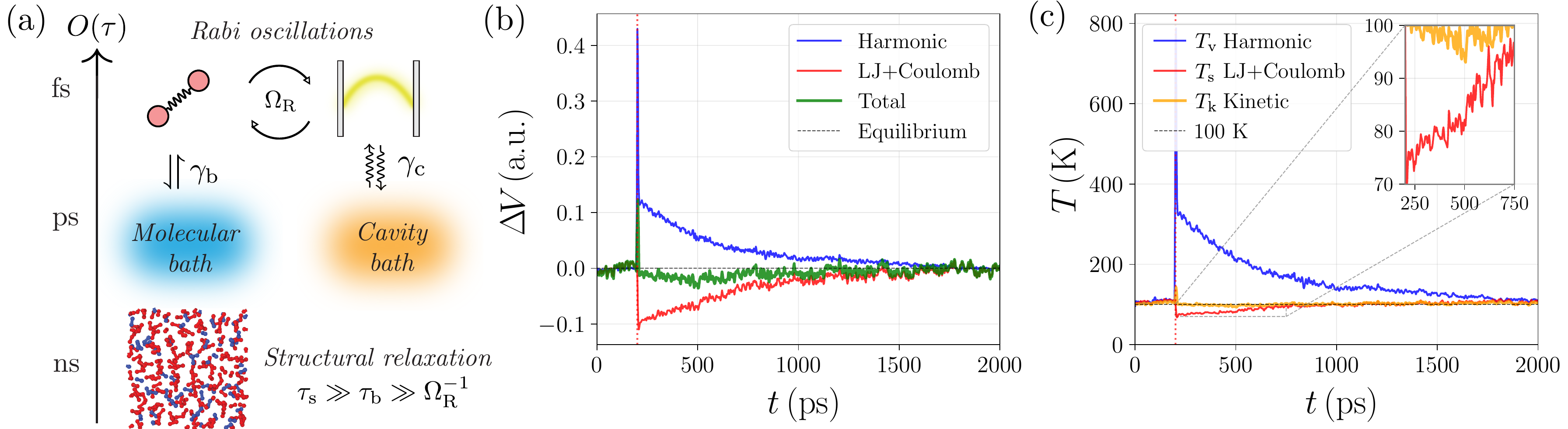}
    \caption{
    Understanding cavity-induced non-thermal aging through energy redistribution and fictive temperatures (for $\lambda = 0.141$~a.u.).
    (a) Hierarchy of kinetic processes: Rabi oscillations ($\Omega_R^{-1}$, femtoseconds), bath dissipation ($\tau_\mathrm{b}$, $\tau_\mathrm{c}$, picoseconds), and structural relaxation ($\tau_\mathrm{s}$, nanoseconds). The clear separation of timescales ensures the cavity selectively excites vibrations without immediately perturbing the slow structural degrees of freedom.
    (b) Potential energy evolution after cavity activation, showing redistribution of energy into vibrational DOFs (blue), which leads to a compensating decrease of intermolecular energy (red), while the total energy (green) remains conserved.
    (c) Fictive temperature evolution: vibrational $T_\mathrm{v}$ (blue), kinetic $T_\mathrm{k}$ (orange), and structural $T_\mathrm{s}$ (red). The structural fictive temperature drops below the bath temperature, confirming cavity-induced non-thermal aging.
    }
    \label{fig:fig3}
\end{figure*}

\section*{Cavity Induced Non-thermal Aging} 

Having confirmed strong light-matter coupling, we now examine how the cavity modifies structural relaxation. 
In conventional aging experiments, an equilibrium supercooled liquid is rapidly quenched to a lower temperature, after which its relaxation time depends explicitly on the elapsed time since the quench, also known as the waiting time $t_\mathrm{w}$. 
The waiting-time dependence of material properties following such a large perturbation in temperature is the hallmark of thermal aging. 
In an optical cavity, we follow a similar protocol in which the temperature change is replaced by the activation of the coupling constant $\lambda(t)$.  
Because structural relaxation in supercooled liquids is intrinsically slow, we expect the cavity can imprint long-lived changes persisting over timescales beyond the equilibrium relaxation time. 
To probe non-thermal aging, we evaluate the intermediate scattering function (ISF), defined as
\begin{equation}
F_{\bm{k}}(t) = \langle \rho_{\bm{k}}(t)\rho^*_{\bm k}(0) \rangle \,, \label{eq:ISF_equil}
\end{equation}
where $\rho_{\bm{k}}(t) = \sum_{i=1}^N e^{\mathrm{i} \bm{k} \cdot \bm{R}_i(t)}$ is the density mode at wavevector $\bm k$.  
$F_{\bm k}(t)$ characterizes the collective molecular motion, and its value at $t=0$ defines the static structure factor $S_{\bm k} = \langle \rho_{\bm{k}}(0)\rho^*_{\bm k}(0) \rangle$, which we use to normalize the ISF as  
\begin{equation}
\phi_{\bm k}(t) := F_{\bm{k}}(t)/S_{\bm k}\,.
\end{equation}
The relaxation time $\tau_\mathrm{s}$ is defined by $\phi_{\bm k}(\tau_\mathrm{s}) = 0.1$, a standard decorrelation threshold; see SI~Sec.~2 for details on observable definitions and averaging.  

\Cref{fig:figure2}a (bottom panel) shows the structural relaxation in the cavity-free case (left) and under increasing light-matter coupling strengths (left to right), evaluated at increasing waiting times (purple to yellow). With increasing coupling strength, the relaxation systematically slows and exhibits hallmarks of aging, where the ISF becomes dependent on the waiting time before returning to its equilibrium behavior. 
Despite the waiting-time dependence, no thermal quench has been applied as the temperature is held fixed at $T=100$~K throughout. 
Instead, the perturbation is the onset of cavity coupling, which slows structural relaxation without altering the bath temperature. 
Thus, we refer to this phenomenon as \textit{cavity-induced non-thermal aging}.

To quantify the cavity-induced contribution to aging, we consider the dimensionless relaxation times, defined as $\tilde{\tau}_\mathrm{s}(t_\mathrm{w}) = \tau_\mathrm{s}(t_\mathrm{w}) / \tau_{\mathrm{s},\lambda=0}(t_\mathrm{w})$, where each relaxation time is normalized to its cavity-free value at the same waiting time, isolating the cavity-induced slowdown from intrinsic thermal aging. \Cref{fig:figure2}b shows the dimensionless relaxation time for different waiting times (purple to yellow) as a function of coupling strength. At short waiting times (purple), the relaxation time increases with coupling strength. It eventually saturates ($\lambda = 0.098$~a.u.)  up to $\sim 4.5\times$ larger relative to its equilibrium value for the larger coupling strengths, which confirms the non-thermal aging observed in \cref{fig:figure2}a and indicates a maximum cavity-induced slowdown. At long waiting times (yellow), the relaxation approaches cavity-free behavior, consistent with the system reaching equilibrium.

The observed slowdown is also long-lived. \Cref{fig:figure2}c replots the dimensionless relaxation time as a function of waiting time $t_\mathrm{w}$ for increasing coupling strengths (blue to red). For finite coupling, the relaxation time remains enhanced over extended waiting times and decays only gradually back to its cavity-free value. The decay timescale increases systematically with coupling strength and saturates at the highest coupling, reaching approximately $2.5$~ns. The system thus retains a memory of the cavity-induced perturbation over nanosecond timescales, far exceeding typical vibrational or polaritonic lifetimes.\cite{dunkelberger2016modified} 
Moreover, the slowdown persists at off-resonant cavity frequencies (SI~Sec.~6.2), underscoring the robustness of the effect.
Strong light-matter coupling thus induces long-lived modifications of structural relaxation.

We now turn to the question of how and why non-thermal aging emerges under cavity coupling. 
Non-thermal aging emerges from two key ingredients: (1) the separation of timescales between the fast light-matter dynamics and slow structural relaxation, and (2) the fact that the system temperature is held fixed so that energetic compensation cannot proceed through heating. 

As shown in \Cref{fig:fig3}a, there are three energy transfer processes with clear hierarchy in timescales:
Rabi oscillations occur on femtosecond timescales ($\Omega_R^{-1}$), energy dissipation through the baths acts on picosecond timescales ($\tau_\mathrm{b}$), and structural relaxation proceeds on nanosecond timescales ($\tau_\mathrm{s}$), such that $\tau_\mathrm{s} \gg \tau_\mathrm{b} \gg \Omega_R^{-1}$.
\Cref{fig:fig3}b shows the evolution of the potential energy after the supercooled liquid is placed in the cavity, i.e., upon instantaneous activation of the coupling. Throughout this process, the temperature is maintained at $T=100$~K, so the kinetic energy remains unchanged. 
Consequently, the energetic response to switching on the cavity must be compensated entirely through the potential energy. 
At initial time, the cavity pumps energy into intramolecular vibrations, creating a spike in vibrational energy (\cref{fig:fig3}b, blue). 
The bath rapidly removes excess energy to maintain the total energy near its equilibrium value. 
To satisfy this effective energy conservation, structural degrees of freedom access lower potential-energy configurations, and both Lennard-Jones and Coulombic contributions decrease simultaneously (\cref{fig:fig3}b, red). 
The system then reaches a deeper potential well from which it must escape, a process considerably slower than equilibrium relaxation.

\paragraph*{Fictive temperatures.} To gain further qualitative understanding of cavity-induced non-thermal aging, we map the potential energy onto a temperature scale using the fictive temperature, the equilibrium temperature at which the system would exhibit the same potential energy as its instantaneous out-of-equilibrium configuration.\cite{dyre2018isomorph,dyre2020isomorph} 
Fictive temperatures can be defined on different components of the potential energy. Given the energy transfer processes responsible for non-thermal aging, we assign a vibrational fictive temperature $T_\mathrm{v}$ (from bond energies $V_\mathrm{b}$) and a structural fictive temperature $T_\mathrm{s}$ (from intermolecular interactions $V_\mathrm{LJ}+V_\mathrm{C}$), which captures the structural degrees of freedom.

We calculate the fictive temperatures by inverting analytical equilibrium potential-energy relations. In equilibrium, the vibrational energy $\varphi(T) := \langle V_\mathrm{b} \rangle$ obeys equipartition at low temperatures, such that $\varphi(T) = N c_\mathrm{v} T$ for $T \to 0$, where $c_\mathrm{v} = \tfrac{1}{4} k_\mathrm{B}$ is the specific heat capacity. Assuming $\varphi(T) \to N v_\infty$ at high $T$, we write the rational interpolant 
\begin{equation}
    \varphi(T) \approx \frac{N c_\mathrm{v} T}{1+(c_\mathrm{v} /v_\infty)T  } \,,  \label{eq:vibenergy}
\end{equation}
and thus the vibrational fictive temperature is
\begin{equation}
T_\mathrm{v}(t) := \varphi^{-1}[V_\mathrm{b}(t)] \,. \label{eq:Tv}
\end{equation}

\begin{figure*}[t]
    \centering
    \includegraphics[width=\linewidth]{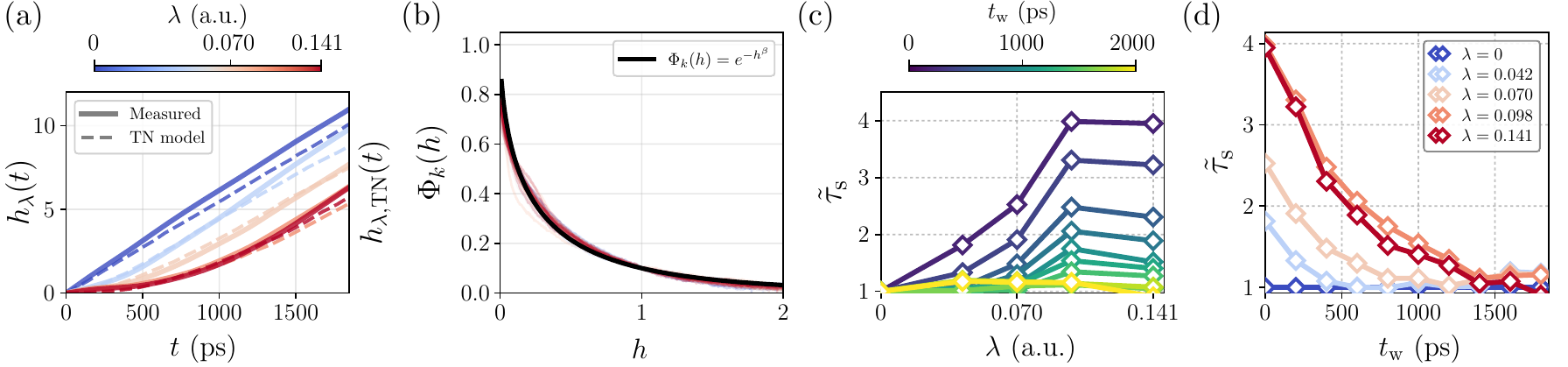}
    \caption{
    Analysis of non-thermal aging through time reparameterization softness.
    (a) Material time $h_\lambda(t)$ reconstructed from ISF measurements (solid lines) with Tool--Narayanaswamy (TN) model predictions (dashed lines) at different $\lambda$ (color-coded).
    The TN model assumes that system relaxation proceeds as if the structure maintains equilibrium at the fictive temperature $T_\mathrm{s}(t)$ at every time $t$. 
    (b) Collapse of all normalized ISFs into a universal stretched exponential $\Phi_{\bm{k}}(h) = e^{-h^\beta}$ with $\beta = 0.55$. The collapse demonstrates that cavity-driven aging obeys the same single-parameter time reparameterization as thermal aging, implying a shared structural relaxation mechanism.
    (c) TN predictions for $h_{\lambda,\mathrm{TN}}(t)$ at different waiting times $t_\mathrm{w}$ (color scale) and (d) at different coupling strengths $\lambda$, showing that the equilibrium-based phenomenological model explains non-thermal aging by capturing the structurally colder state induced by the initial coupling. 
    }
    \label{fig:fig4}
\end{figure*}

For the structural degrees of freedom, we use the equilibrium intramolecular potential energy $\vartheta(T) :=\langle V_\mathrm{LJ}+V_\mathrm{C}\rangle$, whose low-temperature limit follows Rosenfeld-Tarazona scaling, $\vartheta(T) = \vartheta_0+\alpha_0 T^{3/5}$ where $\vartheta_0$ and $\alpha_0$ are constants. \cite{rosenfeld1998density} 
Assuming $\vartheta(T) \to \vartheta_\infty$ at high $T$, we write a similar rational interpolant 
\begin{equation}
\vartheta(T) \approx \vartheta_0+\frac{\alpha_0 T^{3/5}}{1+\frac{\alpha_0}{\vartheta_\infty-\vartheta_0}T^{3/5}} \,. \label{eq:varthetaT}
\end{equation}
Using \cref{eq:varthetaT},  the structural fictive temperature is
\begin{equation}
T_\mathrm{s}(t) := \vartheta^{-1}[V_\mathrm{LJ}(t)+V_\mathrm{C}(t)] \,. \label{eq:ts}
\end{equation}
Since Lennard-Jones and Coulombic interactions govern glassy dynamics, $T_\mathrm{s}$ directly reflects the system’s depth in the potential-energy landscape. Note that all parameters in \cref{eq:vibenergy} and \cref{eq:varthetaT} are obtained by fitting equilibrium potential energies as functions of temperature.

\Cref{fig:fig3}c shows the time evolution of the fictive temperatures for a cavity coupling strength of $\lambda = 0.141$~a.u., including the vibrational fictive temperature $T_\mathrm{v}$ (blue), the structural fictive temperature $T_\mathrm{s}$ (red), and the kinetic temperature (orange). Upon switching on the cavity coupling at $t = 200$~ps, energy is selectively pumped into the vibrational degrees of freedom, leading to a sharp increase of $T_\mathrm{v}$ to approximately $800$~K. By construction, the kinetic temperature remains close to $100$~K at all times, reflecting the fact that the cavity does not heat the system.  In contrast, the structural fictive temperature decreases and approaches $\sim 70$~K, below the initial equilibrium temperature $T = 100$~K. The decrease reflects the compensating response of the intermolecular degrees of freedom to the sudden vibrational energy injection and directly explains the slowdown of structural relaxation at short waiting times. Strong light-matter coupling thus acts as a non-thermal cooling mechanism for the structural subsystem while the bath temperature remains unchanged.

\section*{From Time Reparameterization to a Cooling Protocol}

Having established the mechanism behind non-thermal aging and its description in terms of fictive temperatures, we now show that the long-time dynamics obey time-reparameterization softness (TRS).\cite{Cugliandolo1993,Cugliandolo1997,chamon2002separation,Chamon2002,Chamon2007,Kurchan2023,ghimenti2024cleveralgorithmsglassesdo} 
In glass physics, aging is powerful as it reveals how systems far from equilibrium nevertheless relax along predictable pathways. 
TRS formalizes this idea by showing, when a glassy system is perturbed by a scalar control parameter $\lambda$, e.g., temperature, shear, or here the light-matter coupling, its two-time correlation functions $C_{AA}^{(\lambda)}(t,t^\prime):= \langle A(t) A(t^\prime) \rangle_\lambda$ collapse into a universal functional form.  In particular, 
\begin{equation}
C_{AA}^{(\lambda)}(t,t^\prime)=\mathcal{C}_{AA}\big(h_\lambda(t),h_\lambda(t^\prime)\big) \label{eq:TRS}
\end{equation}
where $h_\lambda(t)$ is a material time that tracks the system’s internal relaxation progress for a given protocol. 
\Cref{eq:TRS} implies that different perturbations do not introduce new relaxation physics but instead rescale the internal clock governing aging. Consequently, the nonequilibrium dynamics induced by the cavity follow the same universal aging trajectories as thermal glasses and can be quantitatively predicted using equilibrium relaxation data via the fictive temperature. This correspondence elevates non-thermal aging from a phenomenological observation to a controlled and predictive tool, laying the foundation for the cooling protocol developed below.

Several approaches exist to show that a glassy system obeys TRS.\cite{ghimenti2024cleveralgorithmsglassesdo,Kurchan2023}
Here, we proceed by invoking material-time translational invariance (MTTI),\cite{bohmer2024time} a specialization of TRS, which further posits that correlations depend only on material-time differences:
\begin{equation}
\mathcal{C}_{AA}(h_\lambda(t),h_\lambda(t^\prime))=\mathcal{\widehat{C}}_{AA}\!\left(h_\lambda(t)-h_\lambda(t^\prime)\right). \label{eq:MTTI}
\end{equation}
where $\mathcal{\widehat{C}}_{AA}(h)$ is a one-variable scalar function. Applying this directly to the ISF, \cref{eq:TRS,eq:MTTI} together yield
\begin{equation}
\phi_{\bm k}^{(\lambda)}(t; t_\mathrm{w}) = \Phi_{\bm k}(h_\lambda(t+t_\mathrm{w})-h_\lambda(t_\mathrm{w})), \label{eq:fkt_trs}
\end{equation}
where $\Phi_{\bm k}(s)$ is the universal scalar function for the ISF and we set the scale $\Phi_{\bm k}(1) = 0.1$. 
To reconstruct $h_\lambda(t)$, we set a series of waiting times $t_\mathrm{w}^{(r)}$ at which we measure the time correlations and compute the structural relaxation time $\tau_\mathrm{s}(t_\mathrm{w}^{(r)},\lambda) \equiv \tau_\mathrm{s}^{(r)}$. 
Inverting \cref{eq:fkt_trs} for every $r$-th waiting time, we obtain the following datapoint:
\begin{equation}
h_\lambda(\tau_\mathrm{s}^{(r)}+t_\mathrm{w}^{(r)}) - h_\lambda(t_\mathrm{w}^{(r)}) =  \Phi_{\bm k}^{-1}(0.1) = 1 \,, \label{eq:const_mtti}
\end{equation}
which we can collectively fit (via least-squares regression) with the initial condition $h_\lambda(0) = 0$ to a parameterization of the material time, $\tilde{h}_\lambda(t)$, as a sum of linear basis functions; see Methods. 

\Cref{fig:fig4}a (solid lines) presents the results of this fitting procedure, revealing a nonlinear growth in $h_\lambda(t)$ that systematically slows with increasing coupling strength, as expected from the cavity-induced non-thermal aging discussed above. However, when all normalized ISFs are replotted using the material time, they collapse into a single master curve (\cref{fig:fig4}b) that empirically follows a stretched exponential form: 
\begin{equation}
\Phi_{\bm k}(h) = e^{-h^\beta} \,, \label{eq:stretchexp}
\end{equation}
where $\beta \approx 0.55$ for all coupling strengths. The collapse confirms that the cavity does not introduce new relaxation physics, but instead provides a controllable, non-thermal way to slow the system’s intrinsic aging dynamics while preserving their universal form, consistent with TRS (\cref{eq:fkt_trs}) and MTTI (\cref{eq:MTTI}).

\begin{figure*}[t]
    \centering
    \includegraphics[width=0.95\linewidth]{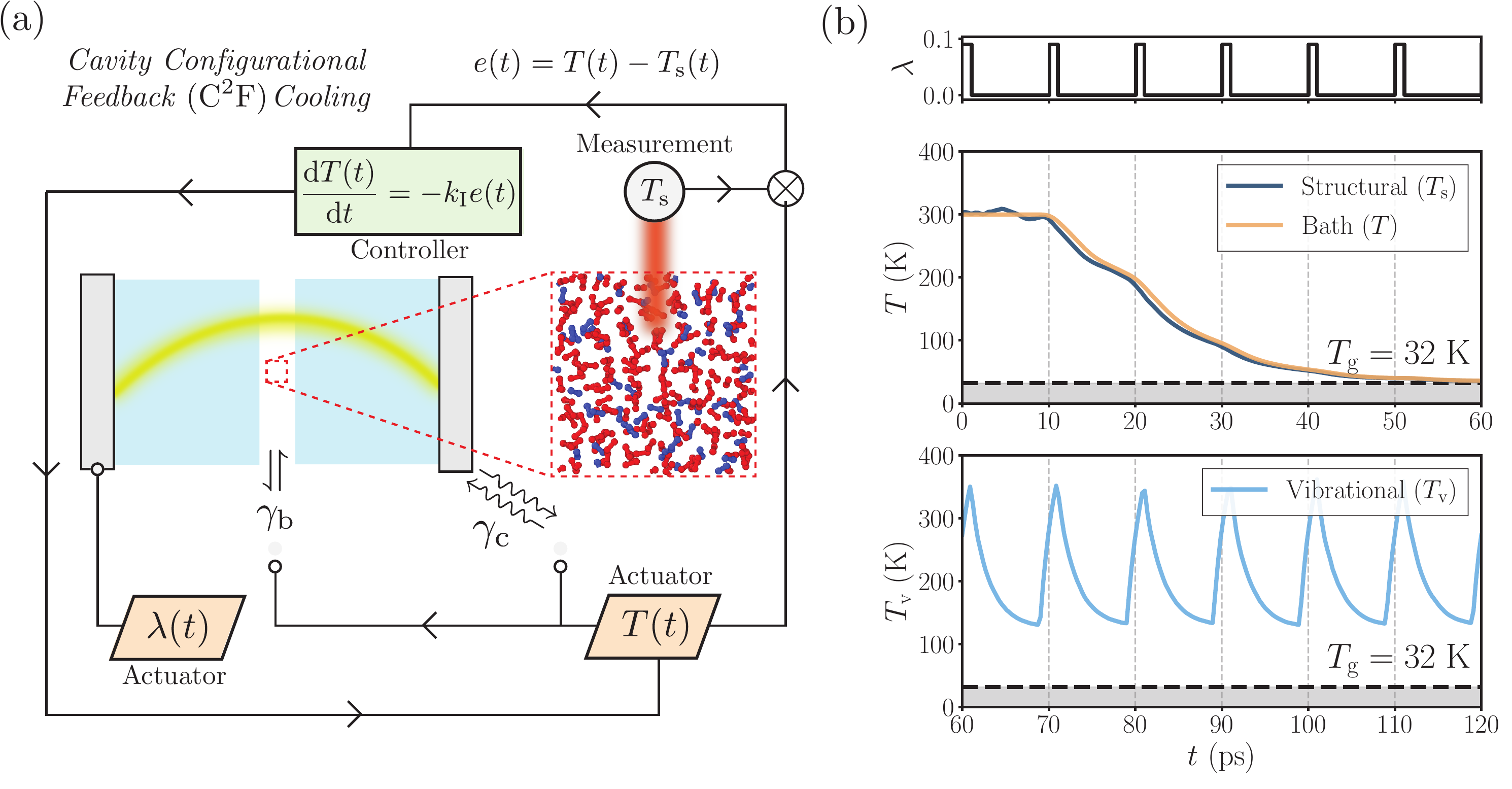}
    \caption{
    Cavity configurational feedback (C$^2$F) cooling applied to a room-temperature liquid.
    (a) Schematic of the C$^2$F protocol. The protocol begins by activating the coupling constant $\lambda(t)$, switching it on over a period, and then turning it off. The fictive temperature is then measured and used by a feedback controller to evaluate the error $e(t)$ between the structural fictive temperature $T_\mathrm{s}$ and the bath temperature $T(t)$. Using the error signal, the controller adjusts the bath temperature  to track structural fictive temperature via closed-loop feedback, preventing the bath from reheating the structure. This step cycles indefinitely until the structure reaches a new low-temperature equilibrium.
    (b) Top: Switching the cavity coupling on and off with a square-wave coupling profile $\lambda(t)$, with coupling peak $\lambda = 0.09$~a.u. Middle: Structural fictive temperature $T_\mathrm{s}$ and bath temperature $T$ converging to $T_\mathrm{g} \approx 32$~K under periodic cavity coupling, demonstrating rapid, cavity-driven cooling from room temperature. By driving the coupling constant at frequencies faster than the slow structural DOFs, the protocol prevents reheating of the system and drive the structure to lower effective temperature. Bottom: synchronized oscillations of $T_\mathrm{v}$, indicating a non-equilibrium steady state in vibrational DOFs consistent with sustained energy exchange.
    }
    \label{fig:fig5}
\end{figure*}

\paragraph*{Correspondence to equilibrium states.}  A key feature of the universal collapse is the inclusion of equilibrium dynamics at zero coupling, where
\begin{equation}
h_{\lambda = 0}(t) = \frac{t}{\tau_\mathrm{s}(T)}.
\end{equation}
When paired with the observed fictive temperatures in \cref{fig:fig3}c, the collapse of aging dynamics into the zero-coupling equilibrium case implies a fundamental correspondence between equilibrium and cavity-driven dynamics.\cite{niss2017mapping,niss2022density} 
Specifically, the cavity-driven system follows the same universal pathways as equilibrium glasses, with its internal clock rescaled by $T_\mathrm{s}(t)$. 

To turn this insight into a quantitative tool, we apply the Tool--Narayanaswamy (TN) model for aging,\cite{Tool1946,Narayanaswamy1971} which asserts that the rate in which material time increases is determined by the instantaneous equilibrium relaxation time:
\begin{equation}
    \frac{\mathrm{d} h_{\lambda,\mathrm{TN}}(t)}{\mathrm{d} t}  = \frac{1}{\tau_\mathrm{s,\mathrm{eq}}[T_\mathrm{s}(t)]} \,, \label{eq:TNmodel}
\end{equation}
where $\tau_\mathrm{s,\mathrm{eq}}[T_\mathrm{s}(t)]$ is the equilibrium structural relaxation time at fictive temperature $T_\mathrm{s}(t)$, obtained from the parabolic-law fit to equilibrium data (SI~Sec.~6.1).

\Cref{fig:fig4}a (dashed lines) shows the material time predicted by the TN model, which qualitatively reproduces $h_\lambda(t)$ (solid lines). From \cref{eq:TNmodel,eq:stretchexp}, we compute the structural relaxation times predicted by the TN model. \Cref{fig:fig4}c shows the TN predictions for the normalized relaxation times $\tilde{\tau}_{\mathrm{s,TN}}$ as a function of $\lambda$ at different waiting times $t_\mathrm{w}$ (color scale); the predictions reproduce the trends observed in \cref{fig:figure2}b. \Cref{fig:fig4}d shows $\tilde{\tau}_\mathrm{s}$ versus $t_\mathrm{w}$ at different coupling strengths, capturing the non-thermal aging and the long-lived memory, consistent with \cref{fig:figure2}c. The TN analysis demonstrates that cavity-induced non-thermal aging is fully determined by equilibrium relaxation properties evaluated at the instantaneous fictive temperature, with the cavity acting solely to rescale the system's internal clock.

\paragraph*{Cavity configurational feedback ($\mathrm{C^2 F}$) cooling.} 

Building on the observation that a liquid coupled to a cavity behaves as if it were structurally colder than its bath temperature, we design a protocol, herein referred to as \textit{cavity configurational feedback ($\mathrm{C^2F}$) cooling}. 
The protocol deliberately exploits non-thermal aging to prepare deeply supercooled liquids and consists of three steps (\cref{fig:fig5}a):

\begin{enumerate}
    \item Measurement of the structural fictive temperature $T_\mathrm{s}(t)$ using \cref{eq:ts}.
    \item Feedback control of the bath temperature $T$, ensuring that $T$ is reduced to match the fictive temperature $T_\mathrm{s}(t)$.  
    \item Periodic modulation of the cavity coupling through controlled on–off switching with specified duty cycle and repetition period.
\end{enumerate}

The core idea of the protocol is as follows (\cref{fig:fig5}a): We couple the supercooled liquid to the cavity ($\lambda = 0.03$~a.u.) for a short, controlled period (\cref{fig:fig5}b, top panel) while keeping the bath temperature fixed. 
As the cavity pumps energy into the fast vibrational modes, the system drives the structural degrees of freedom into deeper potential-energy configurations, thus lowering the structural fictive temperature $T_\mathrm{s}$ and effectively ``cooling" the material. 
We then measure $T_\mathrm{s}$ to reduce the bath temperature accordingly and switch off the cavity. 
During the off-phase, the fast vibrational modes quickly return to equilibrium (\cref{fig:fig5}b, bottom panel), whereas the structural degrees of freedom remain in their deeper configurations due to slower relaxation times. 
We then switch the cavity back on, starting from the new, lower structural state. 
By repeating the cycle of cavity activation and bath adjustment, the system is progressively brought to lower configurational states (\cref{fig:fig5}b, middle panel). 
Eventually, a steady state is reached ($T_g=32$~K) in which the structural degrees of freedom remain equilibrated with the bath ($T \approx T_\mathrm{s}$), thereby avoiding the kinetic arrest that limits conventional cooling.

\Cref{fig:fig5}b shows that, in our simulations, the protocol cools a supercooled liquid from room temperature to $T_\mathrm{g} \approx 32$~K within $\sim 50$~ps. The rapid timescale reflects the controlled feedback and fast cavity modulation accessible and necessary in simulations. The underlying mechanism, however, relies only on the separation of fast vibrational and slow structural dynamics and therefore does not depend on picosecond control. The same principle applies to slower supercooled liquids and experimentally accessible modulation times,\cite{dutta2017alloptical,stone2020optical,yavorskiy2025terahertz} with the material simply evolving on correspondingly longer timescales. Details of the implementation and experimental considerations are provided in Methods, SI~Sec.~4, and the full C$^2$F protocol in SI~Sec.~5.

$\mathrm{C^2F}$ cooling exploits three key ingredients: (i) the ability to periodically control cavity coupling, (ii) the selective and efficient exchange of energy between light and molecular vibrations under strong coupling, and (iii) the intrinsic separation of vibrational and structural timescales, which together establish a new route for controlled manipulation of nonequilibrium dynamics and aging in supercooled liquids.

\section*{Implications and Outlook}

In summary, we show that confining a supercooled liquid inside an optical cavity induces non-thermal aging, i.e., aging without changing the bath temperature. By selectively coupling molecular vibrations with light, the system is driven toward lower-energy basins and thus behaves structurally colder than its surroundings. Crucially, the resulting nonequilibrium dynamics follow the same universal aging behavior as thermally quenched glasses and can be predicted from equilibrium relaxation data through the fictive temperature. Exploiting the effective structural cooling and the timescale separation, we introduced cavity configurational feedback ($\mathrm{C^2F}$) cooling, which uses cavity coupling to access progressively lower structural temperatures. The non-thermal aging mechanism is general and compatible with realistic experiments; considerations including cavity and sample size, achievable coupling strengths, cavity switching and control, multimode effects, experimental access to the structural state, and relevant timescale separation are discussed in SI~Secs.~4 and~6.2.

More broadly, our results establish a new connection between glass physics and strong light--matter interactions. They show that optical cavities can be used not only to modify fast vibrational dynamics, but also to control slow structural relaxation in disordered materials. We expect this work to lay the foundation for light-based control of aging, glass formation, and nonequilibrium materials dynamics, and more generally for accessing regimes that are difficult to reach through purely thermal or mechanical protocols.

\begin{table*}[t]
\centering
\begin{tabular}{lccc}
\toprule
\multicolumn{4}{c}{\textbf{Intramolecular (Harmonic Bonds)}} \\
\midrule
Bond type & $k$ (Hartree/Bohr$^2$) & $R^0$ (Bohr) & $\omega$ (cm$^{-1}$) \\
\midrule
A--A & $0.73204$ & $2.2817$ & $1560$ \\
B--B & $1.4325$ & $2.0744$ & $2433$ \\
\midrule
\multicolumn{4}{c}{} \\
\multicolumn{4}{c}{\textbf{Intermolecular (Lennard-Jones)}} \\
\midrule
Pair type & $\epsilon$ (Hartree) & $\sigma$ (Bohr) & $r_\mathrm{cut}$ (Bohr) \\
\midrule
A--A & $1.6685 \times 10^{-4}$ & $6.2304$ & $15.0$ \\
B--B & $8.3426 \times 10^{-5}$ & $5.4828$ & $15.0$ \\
A--B & $2.5028 \times 10^{-4}$ & $4.9832$ & $15.0$ \\
\multicolumn{4}{l}{Potential is shifted at $r=r_\mathrm{cut}$ to ensure zeroth-order smoothness.} \\
\midrule
\multicolumn{4}{c}{} \\
\multicolumn{4}{c}{\textbf{Electrostatic (Coulombic)}} \\
\midrule
\multicolumn{4}{l}{Partial charges: $\pm 0.3e$ per molecule} \\
\multicolumn{4}{l}{Ewald summation with cut-off radius of 1 nm.} \\
\bottomrule
\end{tabular}
\caption{\label{tab:forcefield}Force field parameters for the molecular Kob-Andersen (mKA) model. Harmonic bond parameters: spring constant $k$ and equilibrium bond length $R^0$. Lennard-Jones parameters: well depth $\epsilon$ and size parameter $\sigma$. Cutoff radius $r_\mathrm{cut}$ applies to all LJ interactions. All quantities in atomic units (Hartree for energy, Bohr for length). Each diatomic molecule has opposing partial charges $\pm 0.3e$ on its two atoms, creating a permanent dipole.}
\end{table*}

\section*{Methods}

\paragraph*{The light-matter Hamiltonian.} Within the dipole approximation and length gauge, the coupled light-matter Hamiltonian reads
\begin{equation}
H = H_\mathrm{M}+H_\mathrm{EM} \,,
\label{eq:tot_hamiltonian}
\end{equation}
where $H_\mathrm{M}$ denotes the molecular Hamiltonian
\begin{equation}
H_\mathrm{M}({ \bm P_i,\bm R_i }) = \sum_{i=1}^{N} \frac{|\bm P_i|^2}{2 M_i} +V({ \bm R_i}) \,,
\end{equation}
with momentum $\bm{P}_i$, position $\bm{R}_i$, mass $M_i$, and potential energy $V({ \bm R_i })$. The single-mode electromagnetic Hamiltonian is\cite{li2020cavity,li2021cavity}
\begin{equation}
H_\mathrm{EM} ({ p_\alpha, q_\alpha }) = \sum_{\alpha=1,2} \left[ \frac{p_\alpha^2}{2}+\frac{\omega_\mathrm{c}^2}{2} \left(q_\alpha +\frac{ \lambda \mu_\alpha}{\omega_\mathrm{c}} \right)^2 \right], \label{eq:Hem}
\end{equation}
with coupling constant $\lambda$, cavity frequency $\omega_c$, and canonical field coordinates $(p_\alpha, q_\alpha)$ along the transverse polarization directions ${\bm{\hat{e}}_\alpha}$. The projected dipole moment is $\mu_\alpha = \bm \mu \cdot \bm e_\alpha$, with $\bm{\mu} = \sum_{i=1}^N z_i e \bm{R}_i$. All degrees of freedom are treated classically; see Refs.\cite{li2020cavity,li2021cavity} and SI~Sec.~1.1 for more details.
 
The total potential energy is $V = V_\mathrm{b}+V_\mathrm{LJ}+V_\mathrm{C}$, where $V_\mathrm{b}$ collects harmonic bond stretches, $V_\mathrm{LJ}$ is the shifted Lennard-Jones potential, and $V_\mathrm{C}$ is the Coulomb interaction. All force-field parameters are listed in \cref{tab:forcefield}.

\paragraph*{Equations of motion and baths.} 
The cavity mode evolves under Langevin dynamics at rate $\gamma_\mathrm{c}$,
\begin{gather}
\dot{q}_{\alpha} = p_\alpha \label{eq:langevin_cavity_1} \,, \\ 
\dot{p}_\alpha = -\omega_\mathrm{c}^2 q_\alpha - \lambda \omega_\mathrm{c} \mu_\alpha -\gamma_\mathrm{c} p_\alpha +\sqrt{2 \gamma_\mathrm{c} k_\mathrm{B} T } \eta_\alpha(t)  \,, \label{eq:langevin_cavity_2}
\end{gather}
where $\eta_\alpha(t)$ is white noise and $k_\mathrm{B}$ is the Boltzmann constant.
The molecular subsystem is coupled to a Bussi–Parrinello thermostat\cite{bussi2007canonical,bussi2008stochastic} at rate $\gamma_\mathrm{b}$, which provides global velocity rescaling while preserving the microscopic trajectory:
\begin{gather}
\dot{\bm{R}}_i = \frac{\bm{P}_i}{M_i} \,, \label{eq:bussi_mol_1} \\
\dot{\bm{P}}_i = -\nabla_{\bm{R}_i} V +\bm F_i^\mathrm{c}  -\gamma_K \bm P_i + \sqrt{2 D_K} \bm P_i \eta_i (t) \,, \label{eq:bussi_mol_2}
\\
\bm F_i^\mathrm{c} = - \sum_{\alpha=1,2} \lambda z_i e \left(\omega_\mathrm{c} q_\alpha + \lambda \mu_\alpha \right)\hat{\bm e}_\alpha \,,
\end{gather}
where $\bm F_i^\mathrm{c}$ is the cavity force, $K$ the total kinetic energy, and $\gamma_K = \frac{1}{2\tau_\mathrm{b}} \left(1-\left(1-\frac{1}{3 N}\right) \frac{\bar{K}}{K}\right)$, $D_K = \gamma_\mathrm{b} \bar{K} / (3 N  K)$, with $\bar{K} = \frac{3}{2}Nk_\mathrm{B} T$. See SI~Sec.~1.2 for the derivation.

\paragraph*{Material-time reconstruction.} \label{app:mattime}
According to material-time translational invariance (MTTI), time-correlation functions across all protocols can be reparameterized as a scalar function of material-time differences alone.
Given a series of waiting times $t_\mathrm{w}^{(r)}$ and corresponding structural relaxation times $\tau_\mathrm{s}^{(r)}$ defined by $\phi_{\bm k}^{(\lambda)}(\tau_\mathrm{s}^{(r)}; t_\mathrm{w}^{(r)}) = 0.1$, we obtain the constraints
\begin{equation}
h_\lambda(t_\mathrm{w}^{(r)} + \tau_\mathrm{s}^{(r)}) - h_\lambda(t_\mathrm{w}^{(r)}) = 1, \quad r = 1, 2, \ldots, N_\mathrm{w}.
\end{equation}
Rather than the iterative construction used in previous works,\cite{bohmer2024time,PedersenDyre2023} we formulate the reconstruction of $h_\lambda(t)$ as a regularized least-squares problem.
We discretize time on a regular grid of $M_\mathrm{grid}$ points, $t_m = m \,\Delta t$ for $m = 0, \ldots, M_\mathrm{grid}-1$, and expand the material time in piecewise linear (hat) basis functions:
\begin{gather}
\tilde{h}_\lambda(t) = \sum_{m=0}^{M_\mathrm{grid}-1} h_{\lambda}^m \,\theta_m(t)
\\
\theta_m(t)= \begin{cases}\frac{t-t_{m-1}}{t_m-t_{m-1}}, & t \in\left[t_{m-1}, t_m\right] \\ \frac{t_{m+1}-t}{t_{m+1}-t_m}, & t \in\left[t_m, t_{m+1}\right] \\ 0, & \text{otherwise,}\end{cases}
\end{gather}
where the coefficients $\boldsymbol{h}_\lambda = [h_{\lambda}^0, \ldots, h_{\lambda}^{M_\mathrm{grid}-1}]^\mathsf{T}$ are the unknowns.
Substituting into the constraints yields the linear system $\mathbf{A}\boldsymbol{h}_\lambda = \boldsymbol{b}$, with $b_r = 1$ and
\begin{equation}
A_{rm} = \theta_m(t_\mathrm{w}^{(r)} + \tau_\mathrm{s}^{(r)}) - \theta_m(t_\mathrm{w}^{(r)}). \label{eq:Amatrix}
\end{equation}
Because the system is overparameterized, we add a Tikhonov smoothness penalty using the tridiagonal second-order finite-difference matrix $\mathbf{D}$ (row entries $[1,\,-2,\,1]$):
\begin{equation}
\boldsymbol{h}_\lambda^* = \arg\min_{\boldsymbol{h}_\lambda} \left\{ \|\mathbf{A}\boldsymbol{h}_\lambda - \boldsymbol{b}\|^2
+ \alpha \|\mathbf{D}\boldsymbol{h}_\lambda\|^2 \right\},
\label{eq:regularized_LS}
\end{equation}
where $\alpha > 0$ controls the smoothness--data tradeoff.
Setting the gradient to zero gives the normal equations
\begin{equation}
\bigl(\mathbf{A}^\mathsf{T}\mathbf{A}
+ \alpha\,\mathbf{D}^\mathsf{T}\mathbf{D}\bigr)\,\boldsymbol{h}_\lambda^*
= \mathbf{A}^\mathsf{T}\boldsymbol{b},
\end{equation}
which are efficiently solved with sparse linear solvers exploiting the banded structure of $\mathbf{D}^\mathsf{T}\mathbf{D}$. See SI~Sec.~2.3 for the full derivation, including the explicit form of $\mathbf{D}$ and the gradient calculation.

\paragraph*{System parameters.}
The cubic simulation box has side length $L = 40$~Bohr with periodic boundary conditions (density $\rho = 0.0078125$).
Coulombic interactions are computed via the PPPM method in \texttt{HOOMD-blue}, and a cell-based neighbor list with buffer distance $\Delta r = 1.0$~Bohr is used, excluding bonded pairs from nonbonded interactions. Both baths operate with time constants $\tau_\mathrm{b} = \tau_\mathrm{c} = 1.0$~ps ($\gamma_\mathrm{b}=\gamma_\mathrm{c}=0.5$~ps$^{-1}$) unless otherwise specified.
The resulting pressure at fixed volume is $\sim 40$--$50$~MPa across the temperatures studied.
Equilibrium structural relaxation data and the parabolic-law parameterization used for the Tool--Narayanaswamy model are reported in SI~Sec.~6.1.

\paragraph*{Simulation protocol.}
Both non-thermal aging and cavity configurational feedback (C$^2$F) cooling follow a three-stage protocol.
In Stage~1, $250$ diatomic molecules are placed on a simple cubic lattice at the target density, with bond lengths sampled from the equilibrium Boltzmann distribution at temperature $T$.
In Stage~2, a single long NVT equilibration run without the cavity mode is performed at $T = 100$ K (non-thermal aging) or $T = 300$ K (C$^2$F cooling); $500$ configurations separated by 300~ps are extracted from the resulting 150~ns trajectory to serve as independent initial conditions.
In Stage~3, each configuration seeds a production run of up to 2.5~ns. A cavity mode is introduced at $t = 0$ with position and momentum drawn from its thermal equilibrium distribution, and the coupling is instantaneously activated at $t_0 = 200$~ps (non-thermal aging) or $t_0 = 20$~ps (C$^2$F cooling).
Energy decompositions are recorded every $0.1$~ps, and the intermediate scattering function $F_{\bm{k}}(t; t_\mathrm{w})$ is computed by averaging over $50$ wavevectors uniformly distributed on a sphere of radius $|\bm k| = 6.0$~a.u., with waiting times $t_\mathrm{w}$ spaced every 200~ps; see SI~Sec.~2 for observable definitions and averaging procedures. The full specification of the C$^2$F protocol, including the feedback controller and square-wave modulation, is given in SI~Sec.~5.

\paragraph*{Numerical methods.}
We use split-operator integration: the cavity mode is propagated via Langevin velocity-Verlet, and molecular degrees of freedom via a two-step velocity-Verlet propagator with stochastic velocity rescaling.\cite{bussi2007canonical,bussi2008stochastic}
Adaptive timestepping adjusts $\Delta t$ based on local truncation error estimates, yielding $\Delta t \sim 1.5$~fs during stable propagation and as low as $\Delta t \sim 10^{-4}$~fs during sudden changes in $\lambda$.
The cavity-molecule coupling force is implemented in both C++ and CUDA within \texttt{cavHOOMD-blue}; benchmarks against the LAMMPS+i-PI reference implementation\cite{li2020cavity,li2021cavity} are presented in SI~Sec.~6.3.
See SI~Sec.~3.1--3.2 for details on the integration schemes and adaptive timestepping.

\section*{Acknowledgments}

This work was supported by the Simons Center for Computational Physical Chemistry (SCCPC) at NYU (SF Grant No. 839534) and the U.S. DOE, Office of Science, BES (Early Career Award No. DE-SC0026328). Additional support was provided in part through NYU IT High Performance Computing resources, services, and staff expertise. Simulations were partially executed on resources supported by the SCCPC. M.R.H. acknowledges support from a postdoctoral fellowship awarded by the SCCPC at NYU.

\bibliography{apssamp}

\providecommand{\noopsort}[1]{}\providecommand{\singleletter}[1]{#1}%
\begin{thebibliography}{50}%
\makeatletter
\providecommand \@ifxundefined [1]{%
 \@ifx{#1\undefined}
}%
\providecommand \@ifnum [1]{%
 \ifnum #1\expandafter \@firstoftwo
 \else \expandafter \@secondoftwo
 \fi
}%
\providecommand \@ifx [1]{%
 \ifx #1\expandafter \@firstoftwo
 \else \expandafter \@secondoftwo
 \fi
}%
\providecommand \natexlab [1]{#1}%
\providecommand \enquote  [1]{``#1''}%
\providecommand \bibnamefont  [1]{#1}%
\providecommand \bibfnamefont [1]{#1}%
\providecommand \citenamefont [1]{#1}%
\providecommand \href@noop [0]{\@secondoftwo}%
\providecommand \href [0]{\begingroup \@sanitize@url \@href}%
\providecommand \@href[1]{\@@startlink{#1}\@@href}%
\providecommand \@@href[1]{\endgroup#1\@@endlink}%
\providecommand \@sanitize@url [0]{\catcode `\\12\catcode `\$12\catcode `\&12\catcode `\#12\catcode `\^12\catcode `\_12\catcode `\%12\relax}%
\providecommand \@@startlink[1]{}%
\providecommand \@@endlink[0]{}%
\providecommand \url  [0]{\begingroup\@sanitize@url \@url }%
\providecommand \@url [1]{\endgroup\@href {#1}{\urlprefix }}%
\providecommand \urlprefix  [0]{URL }%
\providecommand \Eprint [0]{\href }%
\providecommand \doibase [0]{https://doi.org/}%
\providecommand \selectlanguage [0]{\@gobble}%
\providecommand \bibinfo  [0]{\@secondoftwo}%
\providecommand \bibfield  [0]{\@secondoftwo}%
\providecommand \translation [1]{[#1]}%
\providecommand \BibitemOpen [0]{}%
\providecommand \bibitemStop [0]{}%
\providecommand \bibitemNoStop [0]{.\EOS\space}%
\providecommand \EOS [0]{\spacefactor3000\relax}%
\providecommand \BibitemShut  [1]{\csname bibitem#1\endcsname}%
\let\auto@bib@innerbib\@empty
%</preamble>
\bibitem [{\citenamefont {Berthier}\ and\ \citenamefont {Biroli}(2011)}]{berthier2011theoretical}%
  \BibitemOpen
  \bibfield  {author} {\bibinfo {author} {\bibfnamefont {L.}~\bibnamefont {Berthier}}\ and\ \bibinfo {author} {\bibfnamefont {G.}~\bibnamefont {Biroli}},\ }\bibfield  {title} {\bibinfo {title} {Theoretical perspective on the glass transition and amorphous materials},\ }\href {https://doi.org/10.1103/RevModPhys.83.587} {\bibfield  {journal} {\bibinfo  {journal} {Rev. Mod. Phys.}\ }\textbf {\bibinfo {volume} {83}},\ \bibinfo {pages} {587} (\bibinfo {year} {2011})}\BibitemShut {NoStop}%
\bibitem [{\citenamefont {Biroli}\ and\ \citenamefont {Garrahan}(2013)}]{biroli2013perspective}%
  \BibitemOpen
  \bibfield  {author} {\bibinfo {author} {\bibfnamefont {G.}~\bibnamefont {Biroli}}\ and\ \bibinfo {author} {\bibfnamefont {J.~P.}\ \bibnamefont {Garrahan}},\ }\bibfield  {title} {\bibinfo {title} {Perspective: The glass transition},\ }\bibfield  {journal} {\bibinfo  {journal} {J. Chem. Phys.}\ }\textbf {\bibinfo {volume} {138}},\ \href {https://doi.org/10.1063/1.4795539} {10.1063/1.4795539} (\bibinfo {year} {2013})\BibitemShut {NoStop}%
\bibitem [{\citenamefont {Struik}(1978)}]{struik1978physical}%
  \BibitemOpen
  \bibfield  {author} {\bibinfo {author} {\bibfnamefont {L.~C.~E.}\ \bibnamefont {Struik}},\ }\href@noop {} {\emph {\bibinfo {title} {Physical Aging in Amorphous Polymers and Other Materials}}}\ (\bibinfo  {publisher} {Elsevier Scientific Pub. Co.},\ \bibinfo {address} {Amsterdam and New York},\ \bibinfo {year} {1978})\ pp.\ \bibinfo {pages} {xiv, 229}\BibitemShut {NoStop}%
\bibitem [{\citenamefont {Angell}\ \emph {et~al.}(2000)\citenamefont {Angell}, \citenamefont {Ngai}, \citenamefont {McKenna}, \citenamefont {McMillan},\ and\ \citenamefont {Martin}}]{angell2000relaxation}%
  \BibitemOpen
  \bibfield  {author} {\bibinfo {author} {\bibfnamefont {C.~A.}\ \bibnamefont {Angell}}, \bibinfo {author} {\bibfnamefont {K.~L.}\ \bibnamefont {Ngai}}, \bibinfo {author} {\bibfnamefont {G.~B.}\ \bibnamefont {McKenna}}, \bibinfo {author} {\bibfnamefont {P.~F.}\ \bibnamefont {McMillan}},\ and\ \bibinfo {author} {\bibfnamefont {S.~W.}\ \bibnamefont {Martin}},\ }\bibfield  {title} {\bibinfo {title} {Relaxation in glassforming liquids and amorphous solids},\ }\href {https://doi.org/10.1063/1.1286035} {\bibfield  {journal} {\bibinfo  {journal} {J. Appl. Phys.}\ }\textbf {\bibinfo {volume} {88}},\ \bibinfo {pages} {3113} (\bibinfo {year} {2000})}\BibitemShut {NoStop}%
\bibitem [{\citenamefont {Kauzmann}(1948)}]{kauzmann1948nature}%
  \BibitemOpen
  \bibfield  {author} {\bibinfo {author} {\bibfnamefont {W.}~\bibnamefont {Kauzmann}},\ }\bibfield  {title} {\bibinfo {title} {The nature of the glassy state and the behavior of liquids at low temperatures.},\ }\href {https://doi.org/10.1021/cr60135a002} {\bibfield  {journal} {\bibinfo  {journal} {Chem. Rev.}\ }\textbf {\bibinfo {volume} {43}},\ \bibinfo {pages} {219} (\bibinfo {year} {1948})}\BibitemShut {NoStop}%
\bibitem [{\citenamefont {Rodriguez-Tinoco}\ \emph {et~al.}(2022)\citenamefont {Rodriguez-Tinoco}, \citenamefont {Gonzalez-Silveira}, \citenamefont {Ramos},\ and\ \citenamefont {Rodriguez-Viejo}}]{rodriguez2022ultrastable}%
  \BibitemOpen
  \bibfield  {author} {\bibinfo {author} {\bibfnamefont {C.}~\bibnamefont {Rodriguez-Tinoco}}, \bibinfo {author} {\bibfnamefont {M.}~\bibnamefont {Gonzalez-Silveira}}, \bibinfo {author} {\bibfnamefont {M.~A.}\ \bibnamefont {Ramos}},\ and\ \bibinfo {author} {\bibfnamefont {J.}~\bibnamefont {Rodriguez-Viejo}},\ }\bibfield  {title} {\bibinfo {title} {Ultrastable glasses: new perspectives for an old problem},\ }\href {https://doi.org/10.1007/s40766-022-00029-y} {\bibfield  {journal} {\bibinfo  {journal} {Riv. Nuovo Cim.}\ }\textbf {\bibinfo {volume} {45}},\ \bibinfo {pages} {325} (\bibinfo {year} {2022})}\BibitemShut {NoStop}%
\bibitem [{\citenamefont {Ediger}(2017)}]{ediger2017perspective}%
  \BibitemOpen
  \bibfield  {author} {\bibinfo {author} {\bibfnamefont {M.~D.}\ \bibnamefont {Ediger}},\ }\bibfield  {title} {\bibinfo {title} {Perspective: Highly stable vapor-deposited glasses},\ }\bibfield  {journal} {\bibinfo  {journal} {J. Chem. Phys.}\ }\textbf {\bibinfo {volume} {147}},\ \href {https://doi.org/10.1063/1.5006265} {10.1063/1.5006265} (\bibinfo {year} {2017})\BibitemShut {NoStop}%
\bibitem [{\citenamefont {Chen}\ \emph {et~al.}(2023)\citenamefont {Chen}, \citenamefont {Zhao}, \citenamefont {Chi}, \citenamefont {Yan}, \citenamefont {Shen}, \citenamefont {Zou}, \citenamefont {Zhao}, \citenamefont {Liu}, \citenamefont {Yao}, \citenamefont {Zhang} \emph {et~al.}}]{chen2023geological}%
  \BibitemOpen
  \bibfield  {author} {\bibinfo {author} {\bibfnamefont {Z.}~\bibnamefont {Chen}}, \bibinfo {author} {\bibfnamefont {Y.}~\bibnamefont {Zhao}}, \bibinfo {author} {\bibfnamefont {X.}~\bibnamefont {Chi}}, \bibinfo {author} {\bibfnamefont {Y.}~\bibnamefont {Yan}}, \bibinfo {author} {\bibfnamefont {J.}~\bibnamefont {Shen}}, \bibinfo {author} {\bibfnamefont {M.}~\bibnamefont {Zou}}, \bibinfo {author} {\bibfnamefont {S.}~\bibnamefont {Zhao}}, \bibinfo {author} {\bibfnamefont {M.}~\bibnamefont {Liu}}, \bibinfo {author} {\bibfnamefont {W.}~\bibnamefont {Yao}}, \bibinfo {author} {\bibfnamefont {B.}~\bibnamefont {Zhang}}, \emph {et~al.},\ }\bibfield  {title} {\bibinfo {title} {Geological timescales’ aging effects of lunar glasses},\ }\href {https://doi.org/10.1126/sciadv.adi6086} {\bibfield  {journal} {\bibinfo  {journal} {Sci. Adv.}\ }\textbf {\bibinfo {volume} {9}},\ \bibinfo {pages} {eadi6086} (\bibinfo {year} {2023})}\BibitemShut {NoStop}%
\bibitem [{\citenamefont {Sun}\ \emph {et~al.}(2012)\citenamefont {Sun}, \citenamefont {Zhu}, \citenamefont {Wu}, \citenamefont {Cai}, \citenamefont {Gunn},\ and\ \citenamefont {Yu}}]{sun2012stability}%
  \BibitemOpen
  \bibfield  {author} {\bibinfo {author} {\bibfnamefont {Y.}~\bibnamefont {Sun}}, \bibinfo {author} {\bibfnamefont {L.}~\bibnamefont {Zhu}}, \bibinfo {author} {\bibfnamefont {T.}~\bibnamefont {Wu}}, \bibinfo {author} {\bibfnamefont {T.}~\bibnamefont {Cai}}, \bibinfo {author} {\bibfnamefont {E.~M.}\ \bibnamefont {Gunn}},\ and\ \bibinfo {author} {\bibfnamefont {L.}~\bibnamefont {Yu}},\ }\bibfield  {title} {\bibinfo {title} {Stability of amorphous pharmaceutical solids: crystal growth mechanisms and effect of polymer additives},\ }\href {https://doi.org/10.1208/s12248-012-9345-6} {\bibfield  {journal} {\bibinfo  {journal} {AAPS J.}\ }\textbf {\bibinfo {volume} {14}},\ \bibinfo {pages} {380} (\bibinfo {year} {2012})}\BibitemShut {NoStop}%
\bibitem [{\citenamefont {Dragic}\ \emph {et~al.}(2018)\citenamefont {Dragic}, \citenamefont {Cavillon},\ and\ \citenamefont {Ballato}}]{dragic2018materials}%
  \BibitemOpen
  \bibfield  {author} {\bibinfo {author} {\bibfnamefont {P.~D.}\ \bibnamefont {Dragic}}, \bibinfo {author} {\bibfnamefont {M.}~\bibnamefont {Cavillon}},\ and\ \bibinfo {author} {\bibfnamefont {J.}~\bibnamefont {Ballato}},\ }\bibfield  {title} {\bibinfo {title} {Materials for optical fiber lasers: A review},\ }\bibfield  {journal} {\bibinfo  {journal} {Appl. Phys. Rev.}\ }\textbf {\bibinfo {volume} {5}},\ \href {https://doi.org/10.1063/1.5048410} {10.1063/1.5048410} (\bibinfo {year} {2018})\BibitemShut {NoStop}%
\bibitem [{\citenamefont {Lu}\ and\ \citenamefont {Weitz}(2013)}]{lu2013colloidal}%
  \BibitemOpen
  \bibfield  {author} {\bibinfo {author} {\bibfnamefont {P.~J.}\ \bibnamefont {Lu}}\ and\ \bibinfo {author} {\bibfnamefont {D.~A.}\ \bibnamefont {Weitz}},\ }\bibfield  {title} {\bibinfo {title} {Colloidal particles: crystals, glasses, and gels},\ }\href {https://doi.org/10.1146/annurev-conmatphys-030212-184213} {\bibfield  {journal} {\bibinfo  {journal} {Annu. Rev. Condens. Matter Phys.}\ }\textbf {\bibinfo {volume} {4}},\ \bibinfo {pages} {217} (\bibinfo {year} {2013})}\BibitemShut {NoStop}%
\bibitem [{\citenamefont {Vahala}(2003)}]{vahala2003optical}%
  \BibitemOpen
  \bibfield  {author} {\bibinfo {author} {\bibfnamefont {K.~J.}\ \bibnamefont {Vahala}},\ }\bibfield  {title} {\bibinfo {title} {Optical microcavities},\ }\href {https://doi.org/10.1038/nature01939} {\bibfield  {journal} {\bibinfo  {journal} {Nature}\ }\textbf {\bibinfo {volume} {424}},\ \bibinfo {pages} {839} (\bibinfo {year} {2003})}\BibitemShut {NoStop}%
\bibitem [{\citenamefont {Aspelmeyer}\ \emph {et~al.}(2014)\citenamefont {Aspelmeyer}, \citenamefont {Kippenberg},\ and\ \citenamefont {Marquardt}}]{aspelmeyer2014cavity}%
  \BibitemOpen
  \bibfield  {author} {\bibinfo {author} {\bibfnamefont {M.}~\bibnamefont {Aspelmeyer}}, \bibinfo {author} {\bibfnamefont {T.~J.}\ \bibnamefont {Kippenberg}},\ and\ \bibinfo {author} {\bibfnamefont {F.}~\bibnamefont {Marquardt}},\ }\bibfield  {title} {\bibinfo {title} {Cavity optomechanics},\ }\href {https://doi.org/10.1103/RevModPhys.86.1391} {\bibfield  {journal} {\bibinfo  {journal} {Rev. Mod. Phys.}\ }\textbf {\bibinfo {volume} {86}},\ \bibinfo {pages} {1391} (\bibinfo {year} {2014})}\BibitemShut {NoStop}%
\bibitem [{\citenamefont {Cohen-Tannoudji}\ \emph {et~al.}(2024{\natexlab{a}})\citenamefont {Cohen-Tannoudji}, \citenamefont {Dupont-Roc},\ and\ \citenamefont {Grynberg}}]{cohen2024atom}%
  \BibitemOpen
  \bibfield  {author} {\bibinfo {author} {\bibfnamefont {C.}~\bibnamefont {Cohen-Tannoudji}}, \bibinfo {author} {\bibfnamefont {J.}~\bibnamefont {Dupont-Roc}},\ and\ \bibinfo {author} {\bibfnamefont {G.}~\bibnamefont {Grynberg}},\ }\href@noop {} {\emph {\bibinfo {title} {Atom-photon interactions: basic processes and applications}}}\ (\bibinfo  {publisher} {John Wiley \& Sons},\ \bibinfo {year} {2024})\BibitemShut {NoStop}%
\bibitem [{\citenamefont {Cohen-Tannoudji}\ \emph {et~al.}(2024{\natexlab{b}})\citenamefont {Cohen-Tannoudji}, \citenamefont {Dupont-Roc},\ and\ \citenamefont {Grynberg}}]{cohen2024photons}%
  \BibitemOpen
  \bibfield  {author} {\bibinfo {author} {\bibfnamefont {C.}~\bibnamefont {Cohen-Tannoudji}}, \bibinfo {author} {\bibfnamefont {J.}~\bibnamefont {Dupont-Roc}},\ and\ \bibinfo {author} {\bibfnamefont {G.}~\bibnamefont {Grynberg}},\ }\href@noop {} {\emph {\bibinfo {title} {Photons and atoms: introduction to quantum electrodynamics}}}\ (\bibinfo  {publisher} {John Wiley \& Sons},\ \bibinfo {year} {2024})\BibitemShut {NoStop}%
\bibitem [{\citenamefont {Ebbesen}\ \emph {et~al.}(2023)\citenamefont {Ebbesen}, \citenamefont {Rubio},\ and\ \citenamefont {Scholes}}]{ebbesen2023introduction}%
  \BibitemOpen
  \bibfield  {author} {\bibinfo {author} {\bibfnamefont {T.~W.}\ \bibnamefont {Ebbesen}}, \bibinfo {author} {\bibfnamefont {A.}~\bibnamefont {Rubio}},\ and\ \bibinfo {author} {\bibfnamefont {G.~D.}\ \bibnamefont {Scholes}},\ }\bibfield  {title} {\bibinfo {title} {Introduction: polaritonic chemistry},\ }\href {https://doi.org/10.1021/acs.chemrev.3c00637} {\bibfield  {journal} {\bibinfo  {journal} {Chem. Rev.}\ }\textbf {\bibinfo {volume} {123}},\ \bibinfo {pages} {12037} (\bibinfo {year} {2023})}\BibitemShut {NoStop}%
\bibitem [{\citenamefont {Garc\'{i}a-Vidal}\ \emph {et~al.}(2021)\citenamefont {Garc\'{i}a-Vidal}, \citenamefont {Ciuti},\ and\ \citenamefont {Ebbesen}}]{garciavidal2021manipulating}%
  \BibitemOpen
  \bibfield  {author} {\bibinfo {author} {\bibfnamefont {F.~J.}\ \bibnamefont {Garc\'{i}a-Vidal}}, \bibinfo {author} {\bibfnamefont {C.}~\bibnamefont {Ciuti}},\ and\ \bibinfo {author} {\bibfnamefont {T.~W.}\ \bibnamefont {Ebbesen}},\ }\bibfield  {title} {\bibinfo {title} {Manipulating matter by strong coupling to vacuum fields},\ }\href {https://doi.org/10.1126/science.abd0336} {\bibfield  {journal} {\bibinfo  {journal} {Science}\ }\textbf {\bibinfo {volume} {373}},\ \bibinfo {pages} {eabd0336} (\bibinfo {year} {2021})}\BibitemShut {NoStop}%
\bibitem [{\citenamefont {Li}\ \emph {et~al.}(2022)\citenamefont {Li}, \citenamefont {Cui}, \citenamefont {Subotnik},\ and\ \citenamefont {Nitzan}}]{li2022molecular}%
  \BibitemOpen
  \bibfield  {author} {\bibinfo {author} {\bibfnamefont {T.~E.}\ \bibnamefont {Li}}, \bibinfo {author} {\bibfnamefont {B.}~\bibnamefont {Cui}}, \bibinfo {author} {\bibfnamefont {J.~E.}\ \bibnamefont {Subotnik}},\ and\ \bibinfo {author} {\bibfnamefont {A.}~\bibnamefont {Nitzan}},\ }\bibfield  {title} {\bibinfo {title} {Molecular polaritonics: Chemical dynamics under strong light--matter coupling},\ }\href {https://doi.org/10.1146/annurev-physchem-090519-042621} {\bibfield  {journal} {\bibinfo  {journal} {Annu. Rev. Phys. Chem.}\ }\textbf {\bibinfo {volume} {73}},\ \bibinfo {pages} {43} (\bibinfo {year} {2022})}\BibitemShut {NoStop}%
\bibitem [{\citenamefont {Thomas}\ \emph {et~al.}(2016)\citenamefont {Thomas}, \citenamefont {George}, \citenamefont {Shalabney}, \citenamefont {Dryzhakov}, \citenamefont {Varma}, \citenamefont {Moran}, \citenamefont {Chervy}, \citenamefont {Zhong}, \citenamefont {Devaux}, \citenamefont {Genet}, \citenamefont {Hutchison},\ and\ \citenamefont {Ebbesen}}]{thomas2016ground}%
  \BibitemOpen
  \bibfield  {author} {\bibinfo {author} {\bibfnamefont {A.}~\bibnamefont {Thomas}}, \bibinfo {author} {\bibfnamefont {J.}~\bibnamefont {George}}, \bibinfo {author} {\bibfnamefont {A.}~\bibnamefont {Shalabney}}, \bibinfo {author} {\bibfnamefont {M.}~\bibnamefont {Dryzhakov}}, \bibinfo {author} {\bibfnamefont {S.~J.}\ \bibnamefont {Varma}}, \bibinfo {author} {\bibfnamefont {J.}~\bibnamefont {Moran}}, \bibinfo {author} {\bibfnamefont {T.}~\bibnamefont {Chervy}}, \bibinfo {author} {\bibfnamefont {X.}~\bibnamefont {Zhong}}, \bibinfo {author} {\bibfnamefont {E.}~\bibnamefont {Devaux}}, \bibinfo {author} {\bibfnamefont {C.}~\bibnamefont {Genet}}, \bibinfo {author} {\bibfnamefont {J.~A.}\ \bibnamefont {Hutchison}},\ and\ \bibinfo {author} {\bibfnamefont {T.~W.}\ \bibnamefont {Ebbesen}},\ }\bibfield  {title} {\bibinfo {title} {Ground-state chemical reactivity under vibrational coupling to the vacuum electromagnetic field},\ }\href {https://doi.org/10.1002/anie.201605504} {\bibfield  {journal} {\bibinfo  {journal}
  {Angew. Chem. Int. Ed.}\ }\textbf {\bibinfo {volume} {55}},\ \bibinfo {pages} {11462} (\bibinfo {year} {2016})}\BibitemShut {NoStop}%
\bibitem [{\citenamefont {Dunkelberger}\ \emph {et~al.}(2022)\citenamefont {Dunkelberger}, \citenamefont {Simpkins}, \citenamefont {Vurgaftman},\ and\ \citenamefont {Owrutsky}}]{Dunkelberger2022}%
  \BibitemOpen
  \bibfield  {author} {\bibinfo {author} {\bibfnamefont {A.~D.}\ \bibnamefont {Dunkelberger}}, \bibinfo {author} {\bibfnamefont {B.~S.}\ \bibnamefont {Simpkins}}, \bibinfo {author} {\bibfnamefont {I.}~\bibnamefont {Vurgaftman}},\ and\ \bibinfo {author} {\bibfnamefont {J.~C.}\ \bibnamefont {Owrutsky}},\ }\bibfield  {title} {\bibinfo {title} {Vibration-cavity polariton chemistry and dynamics},\ }\href {https://doi.org/10.1146/annurev-physchem-082620-014627} {\bibfield  {journal} {\bibinfo  {journal} {Annu. Rev. Phys. Chem.}\ }\textbf {\bibinfo {volume} {73}},\ \bibinfo {pages} {429} (\bibinfo {year} {2022})}\BibitemShut {NoStop}%
\bibitem [{\citenamefont {Schlawin}\ \emph {et~al.}(2022)\citenamefont {Schlawin}, \citenamefont {Kennes},\ and\ \citenamefont {Sentef}}]{schlawin2022cavity}%
  \BibitemOpen
  \bibfield  {author} {\bibinfo {author} {\bibfnamefont {F.}~\bibnamefont {Schlawin}}, \bibinfo {author} {\bibfnamefont {D.~M.}\ \bibnamefont {Kennes}},\ and\ \bibinfo {author} {\bibfnamefont {M.~A.}\ \bibnamefont {Sentef}},\ }\bibfield  {title} {\bibinfo {title} {Cavity quantum materials},\ }\href {https://doi.org/10.1063/5.0083825} {\bibfield  {journal} {\bibinfo  {journal} {Appl. Phys. Rev.}\ }\textbf {\bibinfo {volume} {9}},\ \bibinfo {pages} {011312} (\bibinfo {year} {2022})}\BibitemShut {NoStop}%
\bibitem [{\citenamefont {Kob}\ and\ \citenamefont {Andersen}(1995)}]{kob1995testing}%
  \BibitemOpen
  \bibfield  {author} {\bibinfo {author} {\bibfnamefont {W.}~\bibnamefont {Kob}}\ and\ \bibinfo {author} {\bibfnamefont {H.~C.}\ \bibnamefont {Andersen}},\ }\bibfield  {title} {\bibinfo {title} {Testing mode-coupling theory for a supercooled binary {Lennard-Jones} mixture {I: The van Hove} correlation function},\ }\href {https://doi.org/10.1103/PhysRevE.51.4626} {\bibfield  {journal} {\bibinfo  {journal} {Phys. Rev. E}\ }\textbf {\bibinfo {volume} {51}},\ \bibinfo {pages} {4626} (\bibinfo {year} {1995})}\BibitemShut {NoStop}%
\bibitem [{\citenamefont {Gaber}\ \emph {et~al.}(2015)\citenamefont {Gaber}, \citenamefont {Takemura}, \citenamefont {Marty}, \citenamefont {Khalil}, \citenamefont {Angelescu}, \citenamefont {Richalot},\ and\ \citenamefont {Bourouina}}]{gaber2015volume}%
  \BibitemOpen
  \bibfield  {author} {\bibinfo {author} {\bibfnamefont {N.}~\bibnamefont {Gaber}}, \bibinfo {author} {\bibfnamefont {Y.}~\bibnamefont {Takemura}}, \bibinfo {author} {\bibfnamefont {F.}~\bibnamefont {Marty}}, \bibinfo {author} {\bibfnamefont {D.~A.~M.}\ \bibnamefont {Khalil}}, \bibinfo {author} {\bibfnamefont {D.}~\bibnamefont {Angelescu}}, \bibinfo {author} {\bibfnamefont {E.}~\bibnamefont {Richalot}},\ and\ \bibinfo {author} {\bibfnamefont {T.}~\bibnamefont {Bourouina}},\ }\bibfield  {title} {\bibinfo {title} {Volume refractometry of liquids using stable optofluidic {F}abry--{P}\'erot resonator with curved surfaces},\ }\href {https://doi.org/10.1117/1.JMM.14.4.045501} {\bibfield  {journal} {\bibinfo  {journal} {J. Micro/Nanolith. MEMS MOEMS}\ }\textbf {\bibinfo {volume} {14}},\ \bibinfo {pages} {045501} (\bibinfo {year} {2015})}\BibitemShut {NoStop}%
\bibitem [{\citenamefont {Li}\ \emph {et~al.}(2020)\citenamefont {Li}, \citenamefont {Subotnik},\ and\ \citenamefont {Nitzan}}]{li2020cavity}%
  \BibitemOpen
  \bibfield  {author} {\bibinfo {author} {\bibfnamefont {T.~E.}\ \bibnamefont {Li}}, \bibinfo {author} {\bibfnamefont {J.~E.}\ \bibnamefont {Subotnik}},\ and\ \bibinfo {author} {\bibfnamefont {A.}~\bibnamefont {Nitzan}},\ }\bibfield  {title} {\bibinfo {title} {Cavity molecular dynamics simulations of liquid water under vibrational ultrastrong coupling},\ }\href {https://doi.org/10.1073/pnas.2009272117} {\bibfield  {journal} {\bibinfo  {journal} {Proc. Natl. Acad. Sci. U.S.A.}\ }\textbf {\bibinfo {volume} {117}},\ \bibinfo {pages} {18324} (\bibinfo {year} {2020})}\BibitemShut {NoStop}%
\bibitem [{\citenamefont {Li}\ \emph {et~al.}(2021)\citenamefont {Li}, \citenamefont {Nitzan},\ and\ \citenamefont {Subotnik}}]{li2021cavity}%
  \BibitemOpen
  \bibfield  {author} {\bibinfo {author} {\bibfnamefont {T.~E.}\ \bibnamefont {Li}}, \bibinfo {author} {\bibfnamefont {A.}~\bibnamefont {Nitzan}},\ and\ \bibinfo {author} {\bibfnamefont {J.~E.}\ \bibnamefont {Subotnik}},\ }\bibfield  {title} {\bibinfo {title} {Cavity molecular dynamics simulations of vibrational polariton-enhanced molecular nonlinear absorption},\ }\href {https://doi.org/10.1063/5.0037623} {\bibfield  {journal} {\bibinfo  {journal} {J. Chem. Phys.}\ }\textbf {\bibinfo {volume} {154}},\ \bibinfo {pages} {094107} (\bibinfo {year} {2021})}\BibitemShut {NoStop}%
\bibitem [{\citenamefont {Bussi}\ \emph {et~al.}(2007)\citenamefont {Bussi}, \citenamefont {Donadio},\ and\ \citenamefont {Parrinello}}]{bussi2007canonical}%
  \BibitemOpen
  \bibfield  {author} {\bibinfo {author} {\bibfnamefont {G.}~\bibnamefont {Bussi}}, \bibinfo {author} {\bibfnamefont {D.}~\bibnamefont {Donadio}},\ and\ \bibinfo {author} {\bibfnamefont {M.}~\bibnamefont {Parrinello}},\ }\bibfield  {title} {\bibinfo {title} {Canonical sampling through velocity rescaling},\ }\href {https://doi.org/10.1063/1.2408420} {\bibfield  {journal} {\bibinfo  {journal} {J. Chem. Phys.}\ }\textbf {\bibinfo {volume} {126}},\ \bibinfo {pages} {014101} (\bibinfo {year} {2007})}\BibitemShut {NoStop}%
\bibitem [{\citenamefont {Bussi}\ and\ \citenamefont {Parrinello}(2008)}]{bussi2008stochastic}%
  \BibitemOpen
  \bibfield  {author} {\bibinfo {author} {\bibfnamefont {G.}~\bibnamefont {Bussi}}\ and\ \bibinfo {author} {\bibfnamefont {M.}~\bibnamefont {Parrinello}},\ }\bibfield  {title} {\bibinfo {title} {Stochastic thermostats: comparison of local and global schemes},\ }\href {https://doi.org/10.1016/j.cpc.2008.01.006} {\bibfield  {journal} {\bibinfo  {journal} {Comput. Phys. Commun.}\ }\textbf {\bibinfo {volume} {179}},\ \bibinfo {pages} {26} (\bibinfo {year} {2008})}\BibitemShut {NoStop}%
\bibitem [{Note1()}]{Note1}%
  \BibitemOpen
  \bibinfo {note} {\protect \url {https://github.com/muhammadhasyim/cav-hoomd}}\BibitemShut {NoStop}%
\bibitem [{\citenamefont {Anderson}\ \emph {et~al.}(2020)\citenamefont {Anderson}, \citenamefont {Glaser},\ and\ \citenamefont {Glotzer}}]{anderson2020hoomd}%
  \BibitemOpen
  \bibfield  {author} {\bibinfo {author} {\bibfnamefont {J.~A.}\ \bibnamefont {Anderson}}, \bibinfo {author} {\bibfnamefont {J.}~\bibnamefont {Glaser}},\ and\ \bibinfo {author} {\bibfnamefont {S.~C.}\ \bibnamefont {Glotzer}},\ }\bibfield  {title} {\bibinfo {title} {{HOOMD-blue: A Python} package for high-performance molecular dynamics and hard particle {Monte Carlo} simulations},\ }\href {https://doi.org/10.1016/j.commatsci.2019.109363} {\bibfield  {journal} {\bibinfo  {journal} {Comput. Mater. Sci.}\ }\textbf {\bibinfo {volume} {173}},\ \bibinfo {pages} {109363} (\bibinfo {year} {2020})}\BibitemShut {NoStop}%
\bibitem [{\citenamefont {Tuckerman}(2023)}]{tuckerman2023statistical}%
  \BibitemOpen
  \bibfield  {author} {\bibinfo {author} {\bibfnamefont {M.~E.}\ \bibnamefont {Tuckerman}},\ }\href@noop {} {\emph {\bibinfo {title} {Statistical mechanics: theory and molecular simulation}}}\ (\bibinfo  {publisher} {Oxford University Press},\ \bibinfo {year} {2023})\BibitemShut {NoStop}%
\bibitem [{\citenamefont {Dunkelberger}\ \emph {et~al.}(2016)\citenamefont {Dunkelberger}, \citenamefont {Spann}, \citenamefont {Fears}, \citenamefont {Simpkins},\ and\ \citenamefont {Owrutsky}}]{dunkelberger2016modified}%
  \BibitemOpen
  \bibfield  {author} {\bibinfo {author} {\bibfnamefont {A.~D.}\ \bibnamefont {Dunkelberger}}, \bibinfo {author} {\bibfnamefont {B.~T.}\ \bibnamefont {Spann}}, \bibinfo {author} {\bibfnamefont {K.~P.}\ \bibnamefont {Fears}}, \bibinfo {author} {\bibfnamefont {B.~S.}\ \bibnamefont {Simpkins}},\ and\ \bibinfo {author} {\bibfnamefont {J.~C.}\ \bibnamefont {Owrutsky}},\ }\bibfield  {title} {\bibinfo {title} {Modified relaxation dynamics and coherent energy exchange in coupled vibration-cavity polaritons},\ }\href {https://doi.org/10.1038/ncomms13504} {\bibfield  {journal} {\bibinfo  {journal} {Nat. Commun.}\ }\textbf {\bibinfo {volume} {7}},\ \bibinfo {pages} {13504} (\bibinfo {year} {2016})}\BibitemShut {NoStop}%
\bibitem [{\citenamefont {Dyre}(2018)}]{dyre2018isomorph}%
  \BibitemOpen
  \bibfield  {author} {\bibinfo {author} {\bibfnamefont {J.~C.}\ \bibnamefont {Dyre}},\ }\bibfield  {title} {\bibinfo {title} {Isomorph theory of physical aging},\ }\bibfield  {journal} {\bibinfo  {journal} {J. Chem. Phys.}\ }\textbf {\bibinfo {volume} {148}},\ \href {https://doi.org/10.1063/1.5022999} {10.1063/1.5022999} (\bibinfo {year} {2018})\BibitemShut {NoStop}%
\bibitem [{\citenamefont {Dyre}(2020)}]{dyre2020isomorph}%
  \BibitemOpen
  \bibfield  {author} {\bibinfo {author} {\bibfnamefont {J.~C.}\ \bibnamefont {Dyre}},\ }\bibfield  {title} {\bibinfo {title} {Isomorph theory beyond thermal equilibrium},\ }\bibfield  {journal} {\bibinfo  {journal} {J. Chem. Phys.}\ }\textbf {\bibinfo {volume} {153}},\ \href {https://doi.org/10.1063/5.0024212} {10.1063/5.0024212} (\bibinfo {year} {2020})\BibitemShut {NoStop}%
\bibitem [{\citenamefont {Rosenfeld}\ and\ \citenamefont {Tarazona}(1998)}]{rosenfeld1998density}%
  \BibitemOpen
  \bibfield  {author} {\bibinfo {author} {\bibfnamefont {Y.}~\bibnamefont {Rosenfeld}}\ and\ \bibinfo {author} {\bibfnamefont {P.}~\bibnamefont {Tarazona}},\ }\bibfield  {title} {\bibinfo {title} {Density functional theory and the asymptotic high density expansion of the free energy of classical solids and fluids},\ }\href {https://doi.org/10.1080/00268979809483145} {\bibfield  {journal} {\bibinfo  {journal} {Mol. Phys.}\ }\textbf {\bibinfo {volume} {95}},\ \bibinfo {pages} {141} (\bibinfo {year} {1998})}\BibitemShut {NoStop}%
\bibitem [{\citenamefont {Cugliandolo}\ and\ \citenamefont {Kurchan}(1993)}]{Cugliandolo1993}%
  \BibitemOpen
  \bibfield  {author} {\bibinfo {author} {\bibfnamefont {L.~F.}\ \bibnamefont {Cugliandolo}}\ and\ \bibinfo {author} {\bibfnamefont {J.}~\bibnamefont {Kurchan}},\ }\bibfield  {title} {\bibinfo {title} {Analytical solution of the off-equilibrium dynamics of a long-range spin-glass model},\ }\href {https://doi.org/10.1103/PhysRevLett.71.173} {\bibfield  {journal} {\bibinfo  {journal} {Phys. Rev. Lett.}\ }\textbf {\bibinfo {volume} {71}},\ \bibinfo {pages} {173} (\bibinfo {year} {1993})}\BibitemShut {NoStop}%
\bibitem [{\citenamefont {Cugliandolo}\ \emph {et~al.}(1997)\citenamefont {Cugliandolo}, \citenamefont {Kurchan}, \citenamefont {Le~Doussal},\ and\ \citenamefont {Peliti}}]{Cugliandolo1997}%
  \BibitemOpen
  \bibfield  {author} {\bibinfo {author} {\bibfnamefont {L.~F.}\ \bibnamefont {Cugliandolo}}, \bibinfo {author} {\bibfnamefont {J.}~\bibnamefont {Kurchan}}, \bibinfo {author} {\bibfnamefont {P.}~\bibnamefont {Le~Doussal}},\ and\ \bibinfo {author} {\bibfnamefont {L.}~\bibnamefont {Peliti}},\ }\bibfield  {title} {\bibinfo {title} {Glassy behaviour in disordered systems with nonrelaxational dynamics},\ }\href {https://doi.org/10.1103/PhysRevLett.78.350} {\bibfield  {journal} {\bibinfo  {journal} {Phys. Rev. Lett.}\ }\textbf {\bibinfo {volume} {78}},\ \bibinfo {pages} {350} (\bibinfo {year} {1997})}\BibitemShut {NoStop}%
\bibitem [{\citenamefont {Chamon}\ \emph {et~al.}(2002{\natexlab{a}})\citenamefont {Chamon}, \citenamefont {Kennett}, \citenamefont {Castillo},\ and\ \citenamefont {Cugliandolo}}]{chamon2002separation}%
  \BibitemOpen
  \bibfield  {author} {\bibinfo {author} {\bibfnamefont {C.}~\bibnamefont {Chamon}}, \bibinfo {author} {\bibfnamefont {M.~P.}\ \bibnamefont {Kennett}}, \bibinfo {author} {\bibfnamefont {H.~E.}\ \bibnamefont {Castillo}},\ and\ \bibinfo {author} {\bibfnamefont {L.~F.}\ \bibnamefont {Cugliandolo}},\ }\bibfield  {title} {\bibinfo {title} {Separation of time scales and reparametrization invariance for aging systems},\ }\href {https://doi.org/10.1103/PhysRevLett.89.217201} {\bibfield  {journal} {\bibinfo  {journal} {Phys. Rev. Lett.}\ }\textbf {\bibinfo {volume} {89}},\ \bibinfo {pages} {217201} (\bibinfo {year} {2002}{\natexlab{a}})}\BibitemShut {NoStop}%
\bibitem [{\citenamefont {Chamon}\ \emph {et~al.}(2002{\natexlab{b}})\citenamefont {Chamon}, \citenamefont {Kennett}, \citenamefont {Castillo},\ and\ \citenamefont {Cugliandolo}}]{Chamon2002}%
  \BibitemOpen
  \bibfield  {author} {\bibinfo {author} {\bibfnamefont {C.}~\bibnamefont {Chamon}}, \bibinfo {author} {\bibfnamefont {M.~P.}\ \bibnamefont {Kennett}}, \bibinfo {author} {\bibfnamefont {H.~E.}\ \bibnamefont {Castillo}},\ and\ \bibinfo {author} {\bibfnamefont {L.~F.}\ \bibnamefont {Cugliandolo}},\ }\bibfield  {title} {\bibinfo {title} {Separation of time scales and reparametrization invariance for aging systems},\ }\href {https://doi.org/10.1103/PhysRevLett.89.217201} {\bibfield  {journal} {\bibinfo  {journal} {Phys. Rev. Lett.}\ }\textbf {\bibinfo {volume} {89}},\ \bibinfo {pages} {217201} (\bibinfo {year} {2002}{\natexlab{b}})}\BibitemShut {NoStop}%
\bibitem [{\citenamefont {Chamon}\ and\ \citenamefont {Cugliandolo}(2007)}]{Chamon2007}%
  \BibitemOpen
  \bibfield  {author} {\bibinfo {author} {\bibfnamefont {C.}~\bibnamefont {Chamon}}\ and\ \bibinfo {author} {\bibfnamefont {L.~F.}\ \bibnamefont {Cugliandolo}},\ }\bibfield  {title} {\bibinfo {title} {Fluctuations in glassy systems},\ }\href {https://doi.org/10.1088/1742-5468/2007/07/P07022} {\bibfield  {journal} {\bibinfo  {journal} {J. Stat. Mech.}\ }\textbf {\bibinfo {volume} {2007}},\ \bibinfo {pages} {P07022} (\bibinfo {year} {2007})}\BibitemShut {NoStop}%
\bibitem [{\citenamefont {Kurchan}(2023)}]{Kurchan2023}%
  \BibitemOpen
  \bibfield  {author} {\bibinfo {author} {\bibfnamefont {J.}~\bibnamefont {Kurchan}},\ }\bibfield  {title} {\bibinfo {title} {Time-reparametrization invariances, multithermalization and the parisi scheme},\ }\href {https://doi.org/10.21468/SciPostPhysCore.6.1.001} {\bibfield  {journal} {\bibinfo  {journal} {SciPost Phys. Core}\ }\textbf {\bibinfo {volume} {6}},\ \bibinfo {pages} {001} (\bibinfo {year} {2023})}\BibitemShut {NoStop}%
\bibitem [{\citenamefont {Ghimenti}\ \emph {et~al.}(2024)\citenamefont {Ghimenti}, \citenamefont {Berthier}, \citenamefont {Kurchan},\ and\ \citenamefont {van Wijland}}]{ghimenti2024cleveralgorithmsglassesdo}%
  \BibitemOpen
  \bibfield  {author} {\bibinfo {author} {\bibfnamefont {F.}~\bibnamefont {Ghimenti}}, \bibinfo {author} {\bibfnamefont {L.}~\bibnamefont {Berthier}}, \bibinfo {author} {\bibfnamefont {J.}~\bibnamefont {Kurchan}},\ and\ \bibinfo {author} {\bibfnamefont {F.}~\bibnamefont {van Wijland}},\ }\bibfield  {title} {\bibinfo {title} {Clever algorithms for glasses work by time reparameterization},\ }\href {https://doi.org/10.1073/pnas.2520818123} {\bibfield  {journal} {\bibinfo  {journal} {Proc. Natl. Acad. Sci. U.S.A.}\ }\textbf {\bibinfo {volume} {121}},\ \bibinfo {pages} {e2520818123} (\bibinfo {year} {2024})}\BibitemShut {NoStop}%
\bibitem [{\citenamefont {B{\"o}hmer}\ \emph {et~al.}(2024)\citenamefont {B{\"o}hmer}, \citenamefont {Gabriel}, \citenamefont {Costigliola}, \citenamefont {Kociok}, \citenamefont {Hecksher}, \citenamefont {Dyre},\ and\ \citenamefont {Blochowicz}}]{bohmer2024time}%
  \BibitemOpen
  \bibfield  {author} {\bibinfo {author} {\bibfnamefont {T.}~\bibnamefont {B{\"o}hmer}}, \bibinfo {author} {\bibfnamefont {J.~P.}\ \bibnamefont {Gabriel}}, \bibinfo {author} {\bibfnamefont {L.}~\bibnamefont {Costigliola}}, \bibinfo {author} {\bibfnamefont {J.-N.}\ \bibnamefont {Kociok}}, \bibinfo {author} {\bibfnamefont {T.}~\bibnamefont {Hecksher}}, \bibinfo {author} {\bibfnamefont {J.~C.}\ \bibnamefont {Dyre}},\ and\ \bibinfo {author} {\bibfnamefont {T.}~\bibnamefont {Blochowicz}},\ }\bibfield  {title} {\bibinfo {title} {Time reversibility during the ageing of materials},\ }\href {https://doi.org/10.1038/s41567-023-02366-z} {\bibfield  {journal} {\bibinfo  {journal} {Nat. Phys.}\ }\textbf {\bibinfo {volume} {20}},\ \bibinfo {pages} {637} (\bibinfo {year} {2024})}\BibitemShut {NoStop}%
\bibitem [{\citenamefont {Niss}(2017)}]{niss2017mapping}%
  \BibitemOpen
  \bibfield  {author} {\bibinfo {author} {\bibfnamefont {K.}~\bibnamefont {Niss}},\ }\bibfield  {title} {\bibinfo {title} {Mapping isobaric aging onto the equilibrium phase diagram},\ }\href {https://doi.org/10.1103/PhysRevLett.119.115703} {\bibfield  {journal} {\bibinfo  {journal} {Phys. Rev. Lett.}\ }\textbf {\bibinfo {volume} {119}},\ \bibinfo {pages} {115703} (\bibinfo {year} {2017})}\BibitemShut {NoStop}%
\bibitem [{\citenamefont {Niss}(2022)}]{niss2022density}%
  \BibitemOpen
  \bibfield  {author} {\bibinfo {author} {\bibfnamefont {K.}~\bibnamefont {Niss}},\ }\bibfield  {title} {\bibinfo {title} {A density scaling conjecture for aging glasses},\ }\href {https://doi.org/10.1063/5.0090869} {\bibfield  {journal} {\bibinfo  {journal} {J. Chem. Phys.}\ }\textbf {\bibinfo {volume} {157}},\ \bibinfo {pages} {054501} (\bibinfo {year} {2022})}\BibitemShut {NoStop}%
\bibitem [{\citenamefont {Tool}(1946)}]{Tool1946}%
  \BibitemOpen
  \bibfield  {author} {\bibinfo {author} {\bibfnamefont {A.~Q.}\ \bibnamefont {Tool}},\ }\bibfield  {title} {\bibinfo {title} {Relation between inelastic deformability and relaxation time in glass},\ }\href {https://doi.org/10.1111/j.1151-2916.1946.tb11592.x} {\bibfield  {journal} {\bibinfo  {journal} {J. Am. Ceram. Soc.}\ }\textbf {\bibinfo {volume} {29}},\ \bibinfo {pages} {240} (\bibinfo {year} {1946})}\BibitemShut {NoStop}%
\bibitem [{\citenamefont {Narayanaswamy}(1971)}]{Narayanaswamy1971}%
  \BibitemOpen
  \bibfield  {author} {\bibinfo {author} {\bibfnamefont {O.~S.}\ \bibnamefont {Narayanaswamy}},\ }\bibfield  {title} {\bibinfo {title} {A model of structural relaxation in glass},\ }\href {https://doi.org/10.1111/j.1151-2916.1971.tb12186.x} {\bibfield  {journal} {\bibinfo  {journal} {J. Am. Ceram. Soc.}\ }\textbf {\bibinfo {volume} {54}},\ \bibinfo {pages} {491} (\bibinfo {year} {1971})}\BibitemShut {NoStop}%
\bibitem [{\citenamefont {Dutta}\ and\ \citenamefont {Rangwala}(2017)}]{dutta2017alloptical}%
  \BibitemOpen
  \bibfield  {author} {\bibinfo {author} {\bibfnamefont {S.}~\bibnamefont {Dutta}}\ and\ \bibinfo {author} {\bibfnamefont {S.~A.}\ \bibnamefont {Rangwala}},\ }\bibfield  {title} {\bibinfo {title} {All-optical switching in a continuously operated and strongly coupled atom-cavity system},\ }\href {https://doi.org/10.1063/1.4978933} {\bibfield  {journal} {\bibinfo  {journal} {Appl. Phys. Lett.}\ }\textbf {\bibinfo {volume} {110}},\ \bibinfo {pages} {121107} (\bibinfo {year} {2017})}\BibitemShut {NoStop}%
\bibitem [{\citenamefont {Stone}\ \emph {et~al.}(2020)\citenamefont {Stone}, \citenamefont {Suleymanzade}, \citenamefont {Taneja}, \citenamefont {Schuster},\ and\ \citenamefont {Simon}}]{stone2020optical}%
  \BibitemOpen
  \bibfield  {author} {\bibinfo {author} {\bibfnamefont {M.}~\bibnamefont {Stone}}, \bibinfo {author} {\bibfnamefont {A.}~\bibnamefont {Suleymanzade}}, \bibinfo {author} {\bibfnamefont {L.}~\bibnamefont {Taneja}}, \bibinfo {author} {\bibfnamefont {D.~I.}\ \bibnamefont {Schuster}},\ and\ \bibinfo {author} {\bibfnamefont {J.}~\bibnamefont {Simon}},\ }\bibfield  {title} {\bibinfo {title} {Optical mode conversion in coupled {Fabry-P{\'e}rot} resonators},\ }\href {https://doi.org/10.1364/OL.400998} {\bibfield  {journal} {\bibinfo  {journal} {Opt. Lett.}\ }\textbf {\bibinfo {volume} {45}},\ \bibinfo {pages} {3988} (\bibinfo {year} {2020})}\BibitemShut {NoStop}%
\bibitem [{\citenamefont {Yavorskiy}\ \emph {et~al.}(2025)\citenamefont {Yavorskiy}, \citenamefont {Suffczy\'nski}, \citenamefont {Kowerdziej}, \citenamefont {Strze\c{z}ysz}, \citenamefont {Wr\'obel}, \citenamefont {Knap},\ and\ \citenamefont {Białek}}]{yavorskiy2025terahertz}%
  \BibitemOpen
  \bibfield  {author} {\bibinfo {author} {\bibfnamefont {D.}~\bibnamefont {Yavorskiy}}, \bibinfo {author} {\bibfnamefont {J.}~\bibnamefont {Suffczy\'nski}}, \bibinfo {author} {\bibfnamefont {R.}~\bibnamefont {Kowerdziej}}, \bibinfo {author} {\bibfnamefont {O.}~\bibnamefont {Strze\c{z}ysz}}, \bibinfo {author} {\bibfnamefont {J.}~\bibnamefont {Wr\'obel}}, \bibinfo {author} {\bibfnamefont {W.}~\bibnamefont {Knap}},\ and\ \bibinfo {author} {\bibfnamefont {M.}~\bibnamefont {Białek}},\ }\bibfield  {title} {\bibinfo {title} {Terahertz magnon-polaritons control using a tunable liquid crystal cavity},\ }\href {https://arxiv.org/abs/2504.11293} {\bibfield  {journal} {\bibinfo  {journal} {arXiv}\ } (\bibinfo {year} {2025})},\ \Eprint {https://arxiv.org/abs/2504.11293} {arXiv:2504.11293} \BibitemShut {NoStop}%
\bibitem [{\citenamefont {Douglass}\ and\ \citenamefont {Dyre}(2022)}]{PedersenDyre2023}%
  \BibitemOpen
  \bibfield  {author} {\bibinfo {author} {\bibfnamefont {I.~M.}\ \bibnamefont {Douglass}}\ and\ \bibinfo {author} {\bibfnamefont {J.~C.}\ \bibnamefont {Dyre}},\ }\bibfield  {title} {\bibinfo {title} {Distance-as-time in physical aging},\ }\href {https://doi.org/10.1103/PhysRevE.106.054615} {\bibfield  {journal} {\bibinfo  {journal} {Phys. Rev. E}\ }\textbf {\bibinfo {volume} {106}},\ \bibinfo {pages} {054615} (\bibinfo {year} {2022})}\BibitemShut {NoStop}%
\end{thebibliography}%

\end{document}